\documentclass[fleqn,usenatbib]{mnras}

\usepackage{newtxtext,newtxmath}
\usepackage[T1]{fontenc}

\DeclareRobustCommand{\VAN}[3]{#2}
\let\VANthebibliography\thebibliography
\def\thebibliography{\DeclareRobustCommand{\VAN}[3]{##3}\VANthebibliography}

\usepackage{graphicx}	% Including figure files
\usepackage{amsmath}	% Advanced maths commands
\usepackage{comment}
\usepackage[dvipsnames]{xcolor}
\usepackage{color}
\usepackage{bm}

\graphicspath{{./picture/}}
\newcommand{\be}{\begin{equation}}
\newcommand{\ee}{\end{equation}}
\newcommand{\bea}{\begin{eqnarray}}
\newcommand{\eea}{\end{eqnarray}}

\newcommand{\bi}{\begin{itemize}}
\newcommand{\ei}{\end{itemize}}

\newcommand{\mH}{m_{\rm H}}

\newcommand{\Hebe}{{\it Hebe} }

\defcitealias{Kiyuna+2023}{K23}

\title[DCBH Formation by Internal Radiative Feedback]{
Internal Radiative Feedback Triggers Global Collapse and Direct-Collapse Black Hole Formation in Metal-free Atomic-Cooling Haloes
}

\author[M. Kiyuna et al.]{
Masaki Kiyuna$^{1}$\thanks{E-mail: kiyuna@astr.tohoku.ac.jp},
Sunmyon Chon$^{1,2,3}$,
Kazuyuki Omukai$^{1}$
and 
Raffaella Schneider$^{3,4,5}$
\\
$^{1}$Astronomical Institute, Graduate School of Science, Tohoku University, Aoba, Sendai 980-8578, Japan\\
$^{2}$Max-Planck-Institut fur Astrophysik, Karl-Schwarzschild-Str. 1, D-85741 Garching, Germany\\
$^{3}$Dipartimento di Fisica, “Sapienza” Università di Roma, Piazzale Aldo Moro 2, 00185 Roma, Italy\\
$^{4}$Sapienza School for Advanced Studies, Viale Regina Elena 291, 00161 Roma, Italy\\
$^{5}$INFN, Sezione Roma1, Dipartimento di Fisica, “Sapienza” Università di Roma, Piazzale Aldo Moro 2, 00185 Roma, Italy
}

\date{Accepted XXX. Received YYY; in original form ZZZ}

\pubyear{2026}

\begin{document}
\label{firstpage}
\pagerange{\pageref{firstpage}--\pageref{lastpage}}
\maketitle

%%%%%%% word limit 250 words (to be checked) %%%%%
\begin{abstract}
The formation of heavy seed black holes from the collapse of
supermassive stars (SMSs) is a promising pathway to the
supermassive black holes (SMBHs).
In the conventional direct-collapse scenario, however, SMS formation
requires an extremely strong external far-ultraviolet (FUV) field to
suppress H$_2$ cooling and fragmentation.
Here, we use cosmological radiation-hydrodynamic simulations to show
that SMSs can form in metal-free atomic-cooling haloes even under
subcritical FUV backgrounds with
$J_{21}=1$--$10^2$.
We follow the evolution for
$\sim1$--$2~{\rm Myr}$
after the onset of star formation, including FUV and ionizing radiation
from stars formed within the halo.
Star formation proceeds through a characteristic two-stage evolution.
H$_2$ cooling initially induces fragmentation and the formation of
ordinary Pop~III stars.
Internal FUV radiation from these stars then photodissociates
H$_2$ and suppresses further fragmentation, while the surrounding
warm gas reservoir contracts under self-gravity and undergoes
global collapse through atomic cooling.
This collapse drives inflow rates of
$\sim0.1$--$1~{\rm M_\odot~yr^{-1}}$
and produces rapidly accreting protostars that reach
$\sim2$--$6\times10^4~{\rm M_\odot}$
by the simulations end.
Since rapid accretion is still ongoing, these objects are expected to
grow to
$\sim10^5~{\rm M_\odot}$
SMSs and then form direct-collapse black holes (DCBHs).
The global collapse is accompanied by clustered star formation,
producing a compact, top-heavy stellar system around the SMSs.
The estimated DCBH abundance,
$n_{\rm DCBH}\sim10^{-3}$--$10^{-2}~{\rm cMpc^{-3}}$,
is high enough to provide a plausible progenitor population for
Little Red Dots and present-day SMBHs.
Our results demonstrate that internal stellar feedback can trigger,
rather than prevent, SMS formation, substantially relaxing the external
FUV requirement for heavy-seed formation.
\end{abstract}


\color{black}
%%%%%%%%%%%%%%%%%%%%%%%%%%%%%%%%%%%%%%%%%%%%%%%%%%

%%%%%%%%%%%%%%%%% BODY OF PAPER %%%%%%%%%%%%%%%%%%
%%%%%%%%%%%%%%%%%%%%%%%%%%%%%%%%%%%%%%%%%%%%%%%%%%%%%%%%%%%%%%%%%%
%%%%%%%%%%%%%%%%%%%%%%%%%%%%%%%%%%%%%%%%%%%%%%%%%%%%%%%%%%%%%%%%%%
%%%%%%%%%%%%%%%%%%%%%%%%%%%%%%%%%%%%%%%%%%%%%%%%%%%%%%%%%%%%%%%%%%

\section{INTRODUCTION}
\label{sec:introduction}

Recent JWST observations have revealed a growing number of candidate
Population III (Pop~III) galaxies with zero or extremely low metallicity
around the epoch of reionization
\citep{Vanzella+2023,Maiolino+2024a,Maiolino+2026,Morishita+2025,Rusta+2026, Nakajima+2026}.
Pop~III stars are generally thought to form in minihaloes with
halo masses of $M_{\rm halo}\sim10^5$--$10^6~{\rm M_\odot}$ at $z\sim20$
\citep{BrommYoshida2011,KlessenGlover2023},
whereas Pop~III galaxies are expected to form in more massive
atomic-cooling haloes with
$M_{\rm halo}\gtrsim10^7$--$10^8~{\rm M_\odot}$
\citep{OhHaiman2002,WiseAbel2007,Greif+2008,BrommYoshida2011}.
The deeper gravitational potentials of these haloes allow them to
retain photoionized gas and sustain star formation despite stellar
feedback, enabling the formation of the first galaxies
\citep{Greif+2008,BrommYoshida2011,Pawlik+2011,Pawlik+2015,
HiranoBromm2017}.
To keep the gas pristine until the halo reaches the atomic-cooling
scale, star formation in its progenitor minihaloes must be suppressed.
FUV irradiation from nearby galaxies or the cosmic background provides
a plausible mechanism for this suppression
\citep{Johnson+2008,Dijkstra+2008, Valiante+2016, Valiante+2018, Trinca+2022}.

Pristine atomic-cooling haloes exposed to FUV radiation are also promising sites for supermassive star (SMS) formation \citep[e.g.,][]{InayoshiVisbalHaiman2020}.
In the standard picture, gas evolution and star formation in these haloes depend strongly on the external FUV intensity.
Under relatively weak FUV irradiation, self-shielding during collapse allows H$_2$ cooling to become effective, leading to conventional Pop~III star formation.
Under an extremely strong FUV field with $J_{21}>J_{\rm crit}\sim10^3$, H$_2$ cooling is suppressed and the gas contracts through atomic cooling at $T\simeq8000~{\rm K}$ \citep{Omukai2001,Shang+2010,Sugimura+2014}.
 \footnote{Here, $J_{21}$ denotes the FUV specific intensity at $h\nu=12.4~{\rm eV}$, the midpoint of the Lyman--Werner (LW) band, in units of $10^{-21}~{\rm erg~s^{-1}~cm^{-2}~Hz^{-1}~sr^{-1}}$.}

The abundance of such SMS formation sites is crucial for understanding
the origin of SMBHs.
SMSs are expected to undergo direct collapse and leave heavy seed BHs
with masses of $\sim10^4$--$10^5~{\rm M_\odot}$, providing massive
seeds for subsequent SMBH growth
\citep{BrommLoeb2003,InayoshiVisbalHaiman2020,ReganVolonteri2024}.
Sensitive near- and mid-infrared observations with JWST have identified
AGN candidates out to $z\sim7$--$10$
\citep{Fan+2023,Maiolino+2024b,Maiolino+2024c}.
At $z\sim5$, JWST has also revealed a large population of compact
objects with broad-line emission, known as Little Red Dots (LRDs)
\citep{Onoue+2023,Matthee+2024,Greene+2024,Kokorev+2024,Akins+2025}.
Although the nature of LRDs remains debated, if they host accreting BHs,
their abundance is orders of magnitude higher than expected from conventional
luminous-quasar luminosity functions.
This may require a mechanism capable of producing heavy BH seeds
relatively frequently in the early Universe.

The conventional DCBH formation scenario, however, requires a pristine atomic-cooling halo to be exposed to an extremely strong external FUV field with $J_{21}>J_{\rm crit}$ \citep{BrommLoeb2003, Visbal+2014b, Regan+2017}.
Such strong irradiation is often accompanied by metal enrichment from the same nearby sources, making haloes that simultaneously satisfy the requirements of intense FUV irradiation and pristine gas relatively rare.
The predicted DCBH number density may therefore be too small to account for the entire SMBH population \citep{Dijkstra+2008,Habouzit+2016,Chon+2016}.

Recently, \citet{ChonOmukai2025} showed that SMSs with $\gtrsim10^5~{\rm M_\odot}$ can form even in slightly metal-enriched gas, provided that $Z\lesssim10^{-3}Z_\odot$.
This relaxes the requirement of pristine gas and substantially broadens the environments in which heavy BH seeds may form.
However, their calculations assume an environment exposed to an extremely strong external FUV radiation field with $J_{21}>J_{\rm crit}$.
It is therefore important to determine whether 
this remaining environmental requirement
can also be relaxed, allowing SMS formation under the more common condition $J_{21}<J_{\rm crit}$.

SMS formation is generally considered difficult under subcritical FUV irradiation, $J_{21}<J_{\rm crit}$.
As the gas density and column density increase during collapse, self-shielding suppresses H$_2$ photodissociation and allows H$_2$ cooling to become effective.
The resulting rapid temperature decrease promotes fragmentation, preventing the nearly monolithic collapse characteristic of the supercritical regime, $J_{21}>J_{\rm crit}$.
Previous numerical simulations have indeed shown that very massive Pop~III stars can form under such conditions, but their masses typically remain $\lesssim10^4~{\rm M_\odot}$
\citep{Regan+2020,Latif+2021}.

Rapid halo growth can dynamically heat the gas and suppress H$_2$ cooling and star formation, allowing the halo to reach the atomic-cooling scale and form an SMS even under subcritical FUV irradiation \citep{Wise+2019}.
This pathway, however, requires H$_2$ cooling and the resulting star formation to be suppressed before SMS formation begins.
Whether a system can instead transition to SMS formation after H$_2$ cooling has already triggered star formation under subcritical FUV irradiation has remained unclear.

Previous simulations of atomic-cooling haloes under subcritical FUV irradiation have mainly emphasized the initial star-formation episode triggered by H$_2$ cooling and the subsequent growth of the resulting massive stars \citep{Regan+2020,Latif+2021}.
Massive Pop~III stars formed during this early phase themselves become strong sources of FUV and ionizing radiation, substantially altering the radiative environment within the halo.
Their FUV radiation can photodissociate H$_2$ in the surrounding gas, suppressing H$_2$ cooling and further fragmentation and potentially allowing the warm gas reservoir to undergo large-scale collapse.
The initial star formation may therefore itself create the conditions for subsequent SMS formation.

The impact of radiation from stars formed earlier in the same halo on subsequent gas collapse has been explored in several studies.
\citet{Dunn+2018} and \citet{Chiaki+2023} suggested that FUV radiation from stars formed earlier in the same halo can suppress H$_2$ cooling and create conditions favourable for SMS formation.
More recently, \citet{Sullivan+2025} found that FUV radiation from stars formed through H$_2$ cooling suppresses H$_2$ cooling and accretion in nearby collapsing clumps, although their calculations did not lead to SMS formation.
These studies demonstrate that radiation from an earlier generation of stars can strongly modify subsequent gas collapse, but the longer-term response of the surrounding halo-scale gas reservoir remains unclear.
In particular, no previous study has self-consistently followed the evolution from the initial star formation triggered by H$_2$ cooling through the subsequent halo-scale collapse to determine whether SMSs actually form and what stellar system emerges.

In this study, we perform cosmological radiation-hydrodynamic simulations of star formation in pristine atomic-cooling haloes exposed to subcritical FUV backgrounds with $J_{21}\lesssim10^2$, following their evolution for $\sim1$--$2~{\rm Myr}$ after the onset of star formation.
We include both FUV and ionizing radiation from Pop~III stars formed within the halo and follow how their radiative feedback affects the subsequent halo-scale gas evolution and star formation.

We find that star formation proceeds through a characteristic two-stage evolution.
Once self-shielding allows H$_2$ cooling to become effective, the gas fragments and Pop~III stars form in the resulting dense cores.
Internal FUV radiation from these stars subsequently photodissociates H$_2$ and suppresses further fragmentation, while the surrounding atomic-cooling gas reservoir continues to contract under self-gravity and eventually undergoes global collapse.
This global collapse produces rapidly accreting SMSs in the central region, surrounded by a compact, strongly top-heavy Pop~III stellar system.
This mechanism substantially relaxes the external FUV intensity required for SMS and heavy BH seed formation, suggesting that heavy seeds may form in a wider range of environments than previously considered.

The remainder of this paper is organized as follows.
In Section~\ref{sec:method}, we describe the numerical framework and simulation setup.
In Section~\ref{sec:results}, we present the star formation and gas evolution within the atomic-cooling halo.
In Section~\ref{sec:discussion}, we discuss heavy seed BH formation and subsequent growth, cosmological abundance, and observational comparisons with Pop~III galaxy candidates and LRDs.
Finally, we summarize our conclusions in Section~\ref{sec:conclusions}.
Throughout this paper, we adopt the cosmological parameters of
\citet{Planck13}.

% --------------------------------------------------
\section{METHODS}
\label{sec:method}

% --------------------------------------------------
\subsection{Numerical Framework}
\label{sec:method_framework}

We perform radiation-hydrodynamic simulations of star formation in an
atomic-cooling halo using the N-body + SPH code GADGET-3.
The simulations follow the coupled evolution of dark matter and gas
under gravity and hydrodynamics, together with radiative feedback from
newly formed stars.
The basic numerical implementation follows \citet{ChonOmukai2025}.

\subsection{Initial Conditions and Simulation Setup}

The comoving box size is $20~h^{-1}~{\rm Mpc}$, and we target an atomic-cooling halo previously identified as a DCBH formation site by \citet{Chon+2016}.
The halo has a spin parameter of $\lambda\simeq0.05$ \citep{Chon+2016}, which is typical of the lognormal distribution of halo spin parameters \citep{Bullock+2001}.
We fix the metallicity at $Z=10^{-6}~Z_\odot$.
Metal-line cooling is negligible at this metallicity,
so thermal evolution is essentially identical to that of primordial gas \citep{Omukai2008}.

Previous simulations of this halo adopted a much stronger external FUV field,
$J_{21}\sim10^3$
\citep{Chon+2016,Chon+2018}.
Here, to examine how heavy seed BH formation depends on the external FUV intensity, we instead consider three lower FUV intensities,
$J_{21}=1,\ 10,\ 10^2$.
The external radiation field is represented by a blackbody spectrum with an effective temperature of
$T_{\rm eff}=10^4~{\rm K}$.
We perform three independent cosmological simulations from the same initial conditions at $z=99$, imposing a different value of $J_{21}$ in each run, because the external FUV intensity affects the chemical, thermal, and dynamical evolution of the gas.
In all runs, the onset of star formation occurs at approximately $z\simeq13$, when the halo mass and virial radius are
$M_{\rm halo}\simeq1.5\times10^8~{\rm M_\odot}$ and
$r_{\rm vir}\simeq0.9~{\rm kpc}$,
respectively.

In the cosmological simulations, gas particles are retained within a physical radius of $1~{\rm kpc}$ around the halo, while dark-matter particles are distributed over a larger region.
When the maximum gas density reaches
$n_{\rm H}>10^{10}~{\rm cm^{-3}}$,
shortly before the onset of star formation, 
we extract a snapshot and cut out the region within $500~{\rm pc}$ of the halo centre.
We use this region as the initial condition for the subsequent non-cosmological simulation, neglecting cosmic expansion.

We define $t=0$ as the onset of star formation in the computational domain and follow the subsequent evolution for up to $t=2~{\rm Myr}$.
Stars with $m_\star\gtrsim10^{2}~{\rm M_\odot}$ have characteristic lifetimes of $\sim2~{\rm Myr}$ \citep{Schaerer2002},
after which feedback from supernova explosions may become important.
In practice, some runs are terminated before $t=2~{\rm Myr}$ because long-lived dense gas structures force the timestep to become extremely short, substantially increasing the computational cost.

% --------------------------------------------------
\subsection{Chemistry, Cooling, and Radiative Transfer}
\label{sec:method_rt}

We solve a primordial non-equilibrium chemical network including
${\rm HI}$, ${\rm HII}$, ${\rm e^-}$,
${\rm H_2}$, and ${\rm H^-}$,
and follow the thermal evolution of the gas including radiative cooling
and thermal processes associated with chemical and photochemical processes.
The reaction rates, cooling and heating functions, and their numerical
implementation follow \citet{Chon+2021} and \citet{ChonOmukai2025}.

The radiation field consists of a uniform external FUV background and local radiation from stars formed within the computational domain.
We use the radiative-transfer implementation described by \citet{ChonLatif2017}.
The radiation field is divided into two energy bins: EUV photons with
$h\nu>13.6~{\rm eV}$ and FUV photons with
$h\nu<13.6~{\rm eV}$.
For radiation from stellar sources, 
we account for the attenuation of the EUV and FUV fluxes during radiative transfer.
The EUV opacity is due to H{\sc i} bound--free absorption, while the FUV opacity includes H$^-$ bound--free absorption, H{\sc i} bound--free absorption from the $2s$ and $2p$ states, H{\sc i} Rayleigh scattering, and Thomson scattering by free electrons.
The attenuated radiation field is then coupled to the thermochemical network through H{\sc i} photoionization, H$_2$ photodissociation, and H$^-$ photodetachment.
The H$_2$ column density is calculated along the radiation path, and
H$_2$ self-shielding is evaluated using the fitting formula of
\citet{WolcottGreen+2011}.

% --------------------------------------------------
\subsection{Particle Splitting and Sink Particle Model}
\label{sec:method_sink}

We employ particle splitting to resolve high-density gas collapse and represent newly formed stars with sink particles.

Particle splitting is implemented following \citet{Kitsionas_Whitworth2002}.
Each SPH particle is split into 13 daughter particles whenever it crosses one of the prescribed density thresholds \citep[see also][]{Chon+2021}.
In both the cosmological and non-cosmological calculations, we adopt
$n_{\rm H}=
10^{1},\,10^{3},\,10^{7},\,
10^{9}~{\rm cm^{-3}}$
as the splitting thresholds.
The initial particle masses are
$120~{\rm M_\odot}$
for dark matter and
$20~{\rm M_\odot}$
for gas.
After four levels of splitting, the gas particle mass reaches
$7\times10^{-4}~{\rm M_\odot}$.

Particle splitting allows us to resolve the Jeans mass with a sufficient number of SPH particles.
At the sink-formation density, 
the Jeans mass is:
\begin{eqnarray}
M_{\rm J}
&\simeq&
0.3~{\rm M_\odot}
\left(
\frac{n_{\rm H}}
{2\times10^{11}~{\rm cm^{-3}}}
\right)^{-1/2}
\left(
\frac{T}{10^3~{\rm K}}
\right)^{3/2},
\end{eqnarray}
and is resolved by
$\gtrsim500$
SPH particles.
This exceeds the
$\sim80$-particle criterion for resolving self-gravitating fragmentation
\citep{BateBurkert1997}.

Cooling is turned off for gas with
$n_{\rm H}>10^{11}~{\rm cm^{-3}}$.
An SPH particle is replaced by a sink particle when its density exceeds
$n_{\rm H}=2\times10^{11}~{\rm cm^{-3}}$
and it lies at a local minimum of the gravitational potential.
The sink radius is set to four times the SPH smoothing length at sink formation.
Estimating the smoothing length from the local Jeans scale gives a characteristic sink radius of
\begin{eqnarray}
r_{\rm sink}
&\sim&
4.5\times10^{-3}~{\rm pc}
\left(\frac{T}{10^4~{\rm K}}\right)^{1/2}.
\end{eqnarray}
SPH particles entering the sink radius are accreted onto the sink,
and their masses are added to the sink-particle mass.
The sink accretion rate is evaluated by averaging the accreted mass
over the dynamical timescale at the sink radius.
Two sink particles are merged when their separation becomes smaller
than the sum of their sink radii.

Radiative feedback is included from all stars represented by sink particles.
Each star is treated as a blackbody radiation source.
Its stellar radius and luminosity are determined from pre-computed protostellar-evolution tables as functions of the stellar mass and mass accretion rate, and the effective temperature is calculated from the resulting luminosity and radius.
The tables are based on one-dimensional protostellar-evolution calculations and cover the normal protostar phase, including pre-main-sequence evolution, as well as the bloated supergiant-protostar configurations associated with high accretion rates
\citep{HosokawaOmukai2009,Hosokawa+2012}.
In our implementation, the stellar properties used for radiative feedback are updated based on the current stellar mass and the mass accretion rate,
which is evaluated from the mass growth over a time interval comparable to the dynamical time at the sink radius.
We therefore do not explicitly follow the gradual Kelvin--Helmholtz contraction and the corresponding increase in
$T_{\rm eff}$
after a high-accretion episode.
We have checked that calculations explicitly including this contraction show no clear quantitative differences in the results discussed here.

After particle splitting, the Jeans mass of a molecular-cooling cloud
core,
$M_{\rm J} \simeq 0.3~{\rm M_\odot}$
is resolved with approximately 500 SPH particles.
This comfortably satisfies the \citet{BateBurkert1997} criterion of $\sim80$ particles for resolving self-gravitating fragmentation.

Fragmentation below the sink scale is not resolved in our simulations.
The resulting multiplicity and stellar mass distribution below
$\sim10$--$100~{\rm M_\odot}$
may therefore depend on numerical resolution.
Throughout this paper, we refer to individual sink particles as stars for simplicity.

\section{RESULTS}
\label{sec:results}

We first provide an overview of the overall evolution in Section~\ref{sec:overview}.
The detailed evolution during each stage is described in Sections~\ref{sec:initial_collapse}--\ref{sec:global_collapse_transition}, followed by the resulting stellar mass distribution in Section~\ref{sec:mass_function}.

% --------------------------------------------------
%%%%% FIG.1 %%%%%
\begin{figure*}
	\centering
    \includegraphics[bb=50 0 800 670,width=15cm,scale=0.2]{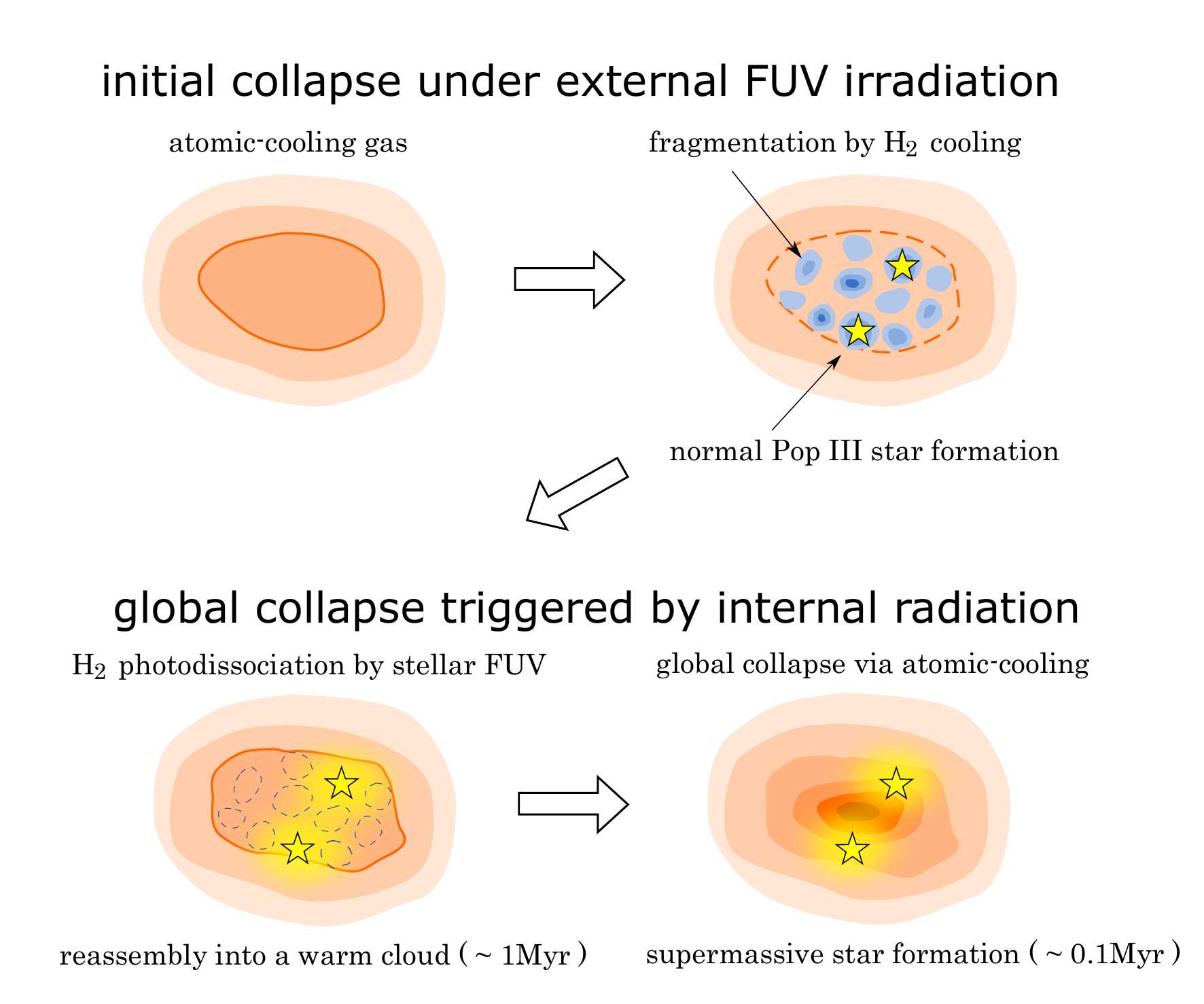}
\caption{
Schematic of the evolutionary sequence of gas collapse in the central region of an atomic-cooling halo exposed to a moderate FUV background.
From left to right, the schematic depicts the initial collapse and fragmentation of the gas, the formation of normal Pop~III stars, the subsequent reassembly of the fragmented gas into a larger warm clump, and the eventual global collapse triggered by internal irradiation.
White arrows indicate the direction of time evolution.
Orange and blue shaded regions represent warm ($T\sim8000~{\rm K}$) and H$_2$-cooled ($T\sim400~{\rm K}$) gas, respectively; within each region, progressively darker shading denotes increasing gas density.
Yellow star symbols indicate normal Pop~III stars.
Dashed outlines mark the preceding spatial configurations of gas structures and are retained as visual guides to illustrate their subsequent fragmentation or reassembly.
}    
\label{fig:schematic_picture}
\end{figure*}

\subsection{Overview: Two modes of star formation during initial and global collapse}
\label{sec:overview}

In all runs, star formation proceeds through a characteristic two-stage evolution.
During the initial collapse, the gas cloud fragments into numerous clumps, with normal Pop~III stars forming in some of them.
Radiative feedback from these stars subsequently suppresses further fragmentation and promotes the reassembly of the fragmented gas into a larger warm cloud.
The subsequent large-scale collapse of this reassembled cloud, which we refer to as global collapse, drives rapid star formation and the growth of SMSs.
While the reassembly of the gas occurs over $\sim1~{\rm Myr}$, the subsequent runaway collapse from $n_{\rm H}\sim10^5~{\rm cm^{-3}}$ proceeds over only $\sim0.1~{\rm Myr}$.
Figure~\ref{fig:schematic_picture} schematically summarizes this evolutionary sequence inferred from the simulations.
The thermochemical and dynamical evolution during each stage is described in detail in Sections~\ref{sec:initial_collapse}--\ref{sec:global_collapse_transition}.

\begin{figure}
    \includegraphics[bb=20 30 400 330,width=10cm,scale=0.2]{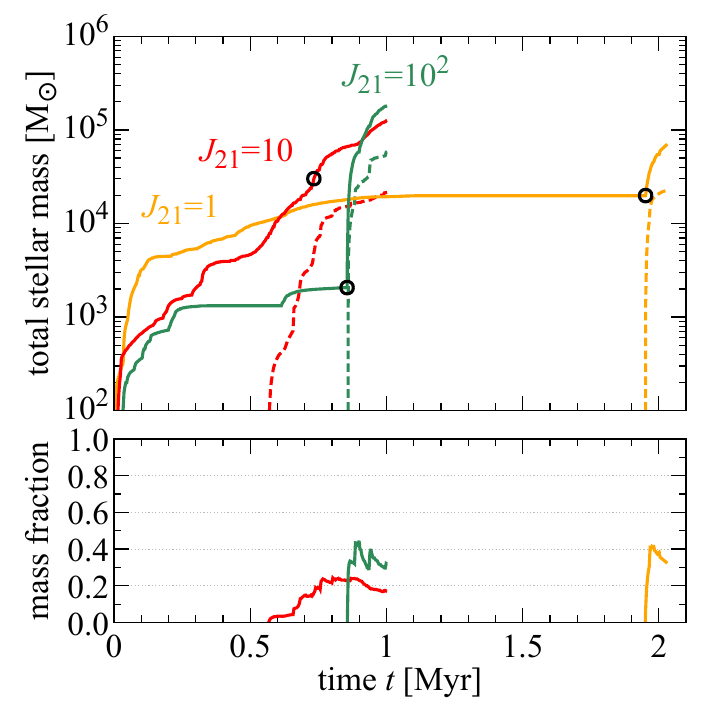}
\caption{
Time evolution of stellar masses in the atomic-cooling halo in the three runs.
The horizontal axis shows the elapsed time since the onset of the initial star formation, $t$.
Hereafter, the ``most massive star'' refers to the star particle that is the most massive
at the end of each simulation.
\textit{Top panel:} Evolution of the total stellar mass (solid lines) and the mass of the
most massive star (dashed lines).
The green, red, and orange lines correspond to the runs with
$J_{21}=10^2$, $10$, and $1$, respectively.
Black open circles mark the onset of 
star formation associated with the
global collapse (see text for definition).
\textit{Bottom panel:} Fraction of the total stellar mass contained in the most massive star.
}
\label{fig:t-Mtot}
\end{figure}

Figure~\ref{fig:t-Mtot} 
shows the stellar mass evolution in each run.
The upper panel shows the time evolution of the total stellar mass.
In all runs, the initial episode of star formation is followed by a period of relatively weak star-formation activity.
The total stellar mass then begins to increase rapidly around the epochs marked by the
black open circles.
These epochs are identified as the onset of star formation induced by the global collapse, as examined in detail in Section~\ref{sec:global_collapse_transition}, and mark the transition to the subsequent SMS formation phase.
The increase is more gradual in the $J_{21}=10$ run, where substantial star formation already occurs before the global collapse.

Star formation associated with the global collapse begins at
$t=0.86~{\rm Myr}$ for $J_{21}=10^2$,
$t=0.73~{\rm Myr}$ for $J_{21}=10$,
and $t=1.95~{\rm Myr}$ for $J_{21}=1$.
By the end of each simulation, the most massive star has grown to
more than $10^4~{\rm M_\odot}$ and contains approximately $20$--$40~\%$
of the total stellar mass (see the bottom panel of Figure~\ref{fig:t-Mtot}).
Thus, a substantial fraction of the stellar mass formed during the global-collapse
phase is concentrated in the most massive star.
We terminated the simulations at $t=1~{\rm Myr}$ for the $J_{21}=10^2$ and $10$ runs and at $t=2~{\rm Myr}$ for the $J_{21}=1$ run because the increasingly short timesteps made further evolution computationally expensive.

\begin{figure*}
	\centering
    \includegraphics[bb=20 0 500 315,width=15cm,scale=0.2]{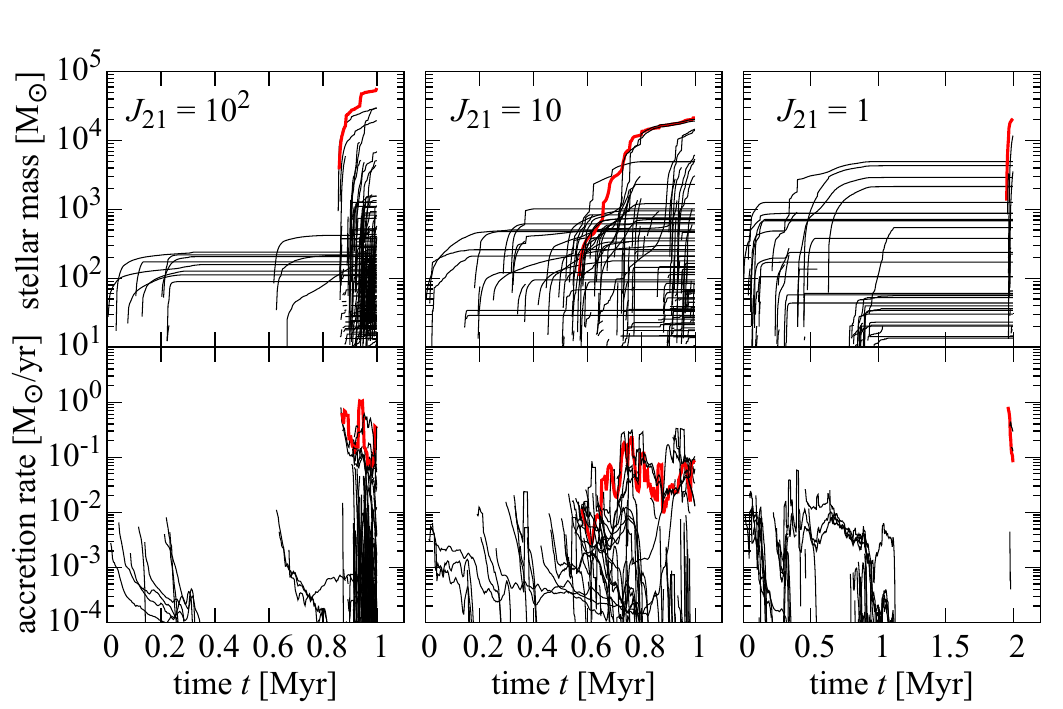}
\caption{
Time evolution of individual stellar masses and their accretion rates in the atomic-cooling halo for the three runs.
The horizontal axis shows the elapsed time since the onset of the initial star formation, $t$, while the vertical axis represents either the stellar mass or the mass-accretion rate.
\textit{Top panels:} Mass evolution of individual stars.
Black lines show all star particles, while the red line highlights the star particle that becomes the most massive by the end of each simulation.
From left to right, the panels correspond to the runs with
$J_{21}=10^2$, $10$, and $1$.
\textit{Bottom panels:} Time evolution of the mass-accretion rates of individual stars.
Black lines show all star particles, while the red line highlights the same star particle as in the corresponding top panel.
}    
\label{fig:t-Msinks}
\end{figure*}

Figure~\ref{fig:t-Msinks} shows the mass evolution and mass-accretion rates of individual star particles.
The upper panel shows that stars formed during the early phase generally cease growing at $m_\star\lesssim10^2$--$10^3~{\rm M_\odot}$, 
although a few reach masses above $10^3~{\rm M_\odot}$.
In contrast, a small number of stars newly formed during the global-collapse phase undergo sustained mass accretion and grow into SMSs.
Numerous lower-mass stars are also formed during this phase.
Thus, SMS formation in our simulations occurs together with the formation of a surrounding stellar cluster.

The total stellar mass and the mass of the most massive star at the end of each simulation are summarized in Table~\ref{tab:summary}.
For example, in the $J_{21}=10^2$ run, the total stellar mass reaches $\sim1.8\times10^5~{\rm M_\odot}$, while the most massive star grows to $\sim5.8\times10^4~{\rm M_\odot}$.
The lower panel of Figure~\ref{fig:t-Msinks} shows that the most massive stars maintain high accretion rates of
$\dot{m}_\star\sim10^{-1}~{\rm M_\odot~yr^{-1}}$,
which, if sustained, would allow them to grow into SMSs with
$m_\star\sim10^5~{\rm M_\odot}$ on a timescale of $\sim1~{\rm Myr}$.

The high accretion rates onto the most massive stars persist until the end of the simulations, suggesting that these stars may continue to grow substantially beyond the masses reached in the present calculations.
This further growth may be affected by supernova explosions from the earlier stellar population, which could modify the subsequent gas supply.
We discuss this uncertainty in Section~\ref{sec:implications}.

Although the three runs differ in their detailed time evolution and degree of fragmentation, they share the same broad evolutionary sequence:
the formation of normal Pop~III stars during the initial collapse,
the reassembly of the fragmented gas into a larger warm cloud,
and the subsequent formation of SMSs during the global collapse.

%------------------------------------------------------------------------%
\begin{table}
\centering
\caption{
Summary of the simulations and resulting stellar properties.
For each run, we list the external FUV intensity, the epoch of the global collapse, the simulation end time, the total stellar mass at the final snapshot, and the mass of the most massive star particle.
}
\label{tab:summary}
\begin{tabular}{cccccc}
\hline
Run &
$J_{21}$ &
$t_{\rm gc}$ &
$t_{\rm end}$ &
$M_{\star,\rm tot}$ &
$m_{\star,\rm max}$ \\
&
&
$[{\rm Myr}]$ &
$[{\rm Myr}]$ &
$[{\rm M_\odot}]$ &
$[{\rm M_\odot}]$ \\
\hline

J1e2 & $10^2$ & 0.86 & 1 & $1.8\times 10^5$ & $5.8\times 10^4$ \\
J1e1 & 10     & 0.73 & 1 & $1.3\times 10^5$ & $2.2\times 10^4$ \\
J1e0 & 1      & 1.95 & 2 & $7.1\times 10^4$ & $2.3\times 10^4$ \\

\hline
\end{tabular}
\end{table}
%------------------------------------------------------------------------%
%------------------------------------------------------------------------%
%%%%% FIG.1 %%%%%
%------------------------------------------------------------------------%
%------------------------------------------------------------------------%
%%%%% FIG.2 %%%%%
%------------------------------------------------------------------------%
\subsection{
Initial Collapse: Pop III Star Formation via H$_2$ Cooling}
\label{sec:initial_collapse}

\begin{figure}
	\centering
    \includegraphics[bb=0 15 400 370,width=15cm,scale=0.2]{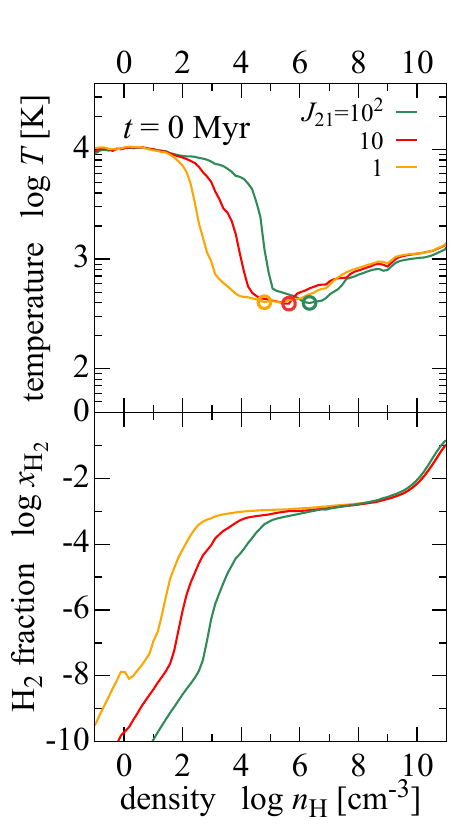}
\caption{
Temperature (top) and H$_2$ fraction (bottom) as functions of density just before the formation of the first protostar in the atomic-cooling halo for the three runs.
The plotted values are mass-weighted averages over gas particles in each density bin.
The green, red, and orange lines correspond to the runs with $J_{21}=10^2$, $10$, and $1$, respectively.
Black open circles in the top panel mark the temperature minimum along each curve.
}
\label{fig:rho-T_first}
\end{figure}

We first examine star formation during the initial collapse.
Figure~\ref{fig:rho-T_first} shows the gas temperature and H$_2$ fraction as functions of density just before the formation of the first protostar in the three runs.
The temperature and H$_2$ fraction are evaluated as mass-weighted mean values in each density bin.
For all external FUV intensities considered, the gas temperature drops sharply from $\sim10^4~{\rm K}$, characteristic of atomic cooling, at $n_{\rm H}\sim10^2$--$10^4~{\rm cm^{-3}}$.
As the central gas becomes denser, H$_2$ self-shielding against the external FUV radiation becomes effective, allowing the H$_2$ fraction to increase and H$_2$ cooling to lower the gas temperature.
This transition shifts to higher densities with increasing external FUV intensity.
The H$_2$ fraction saturates at $\sim10^{-3}$, as expected for H$_2$ formation through the H$^-$ channel.
The subsequent thermal evolution follows the standard H$_2$-cooling track of primordial star formation
\citep{Omukai2000, Omukai2001}.
The transition from the initial atomic-cooling track to the molecular-cooling track is consistent with previous studies of primordial collapse under sub-critical external FUV fields with $J_{21}<J_{\rm crit}$ \citep[e.g.][]{Shang+2010,Sugimura+2014}.

Figure~\ref{fig:rho-T_splash_J1e2} shows the gas structure within the central few tens of parsecs for the $J_{21}=10^2$ run.
At the onset of initial star formation, a gas structure with a characteristic density of $\sim10^3~{\rm cm^{-3}}$ extends over the central $\sim10~{\rm pc}$ region and contains numerous dense cores reaching $\gtrsim10^4~{\rm cm^{-3}}$.
These dense cores have lower temperatures and higher H$_2$ fractions than the surrounding gas, with H$_2$ abundances sufficiently high for efficient H$_2$ cooling.
The individual cores have characteristic sizes of $\sim1~{\rm pc}$, and approximately $10^2$ such cores are distributed throughout the $\sim10~{\rm pc}$ region.

This fragmentation is triggered by the rapid temperature decrease associated with the onset of H$_2$ cooling, 
which reduces thermal pressure support against self-gravity.
The characteristic densities at which the cores form, $n_{\rm H}\sim10^{4-5}~{\rm cm^{-3}}$, approximately coincide with those at which the temperature drops owing to H$_2$ cooling in Figure~\ref{fig:rho-T_first}.
For a dense core with $n_{\rm H}\sim10^4~{\rm cm^{-3}}$ and
$T\sim4\times10^2~{\rm K}$, the Jeans length is
\begin{eqnarray}
\lambda_{\rm J}
&\sim&
1~{\rm pc}
\left(\frac{n_{\rm H}}{10^4~{\rm cm^{-3}}}\right)^{-1/2}
\left(\frac{T}{4\times10^2~{\rm K}}\right)^{1/2},
\end{eqnarray}
which is comparable to the sizes of the individual cores seen in
Figure~\ref{fig:rho-T_splash_J1e2}.
On a larger scale, the overall $\sim10~{\rm pc}$ extent of the core
distribution is comparable to the Jeans length of the surrounding warm gas
with $n_{\rm H}\sim10^3~{\rm cm^{-3}}$ and
$T\sim8\times10^3~{\rm K}$.
The ratio of the Jeans mass of this warm atomic gas to that of the
H$_2$-cooled gas is
\begin{eqnarray}
\frac{M_{\rm J}(10^3~{\rm cm^{-3}}, 8\times10^3~{\rm K})}
{M_{\rm J}(10^4~{\rm cm^{-3}}, 4\times10^2~{\rm K})}
\sim10^2.
\end{eqnarray}
Interestingly, this ratio is of the same order as the number of dense cores present at the onset of initial star formation in Figure~\ref{fig:rho-T_splash_J1e2}.
The initial cloud structure can therefore be interpreted as a self-gravitating warm gas structure on a scale of $\sim10~{\rm pc}$ that fragments through H$_2$ cooling into numerous pc-scale cores.
Similar fragmentation is also seen in the $J_{21}=10$ and $1$ runs shown in Figures~\ref{fig:rho-T_splash_J1e1} and
\ref{fig:rho-T_splash}, although the detailed morphology differs among
the runs.
Note that
$J_{21}=10^2$ maps use a smaller field of view than those for the $J_{21}=10$ and $1$ runs to better show the more spatially confined fragmenting region.

We next estimate how the characteristic fragmentation mass varies with the external FUV intensity.
The characteristic mass scale can be estimated from the Jeans mass at the temperature minimum in Figure~\ref{fig:rho-T_first}.
A stronger external FUV field requires a larger column density for effective H$_2$ self-shielding, shifting the onset of the temperature decline and the subsequent temperature minimum to higher densities.
Accordingly, the characteristic Jeans mass increases toward weaker external FUV fields, from 
$\sim 6\times 10^2~{\rm M}_\odot$
for $J_{21}=10^2$ to $\sim2\times 10^3~{\rm M_\odot}$ for $J_{21}=1$.
The relation between this fragmentation mass scale and the stellar mass distribution that subsequently develops is discussed in
Section~\ref{sec:mass_function}.

%------------------------------------------------------------------------%
%%%%% FIG.4 %%%%%
%------------------------------------------------------------------------%

\subsection{Suppression of Fragmentation by Internal Radiative Feedback}
\label{sec:radiative_feedback}

After the onset of star formation, radiative feedback from the newly formed stars rapidly alters the small-scale gas structure.
The pc-scale H$_2$-cooled clumps formed during the initial collapse gradually disappear, while further small-scale fragmentation and star formation are strongly suppressed.
At the same time, the larger self-gravitating gas structure is not dispersed, but instead reassembles into a warmer and more coherent cloud.
As shown in the following section, this reassembled warm cloud subsequently undergoes global collapse.

At the onset of star formation, $\sim10^2$ 
dense cores produced by H$_2$ cooling are distributed within a larger, self-gravitating gas structure extending over $\sim10~{\rm pc}$.
In the $J_{21}=10^2$ run, however, only seven stars have formed by $t=0.5~{\rm Myr}$.
Thus, most of the dense cores produced during the initial collapse do not subsequently proceed to star formation.
Radiative feedback from the newly formed stars, through photodissociation, reduces the H$_2$ fraction in the central region.
Consequently, the low-temperature, H$_2$-rich clumpy structure present at the onset of star formation largely disappears by $t=0.5~{\rm Myr}$.
In contrast, the larger central gas structure on a scale of
$\sim10~{\rm pc}$ remains largely intact.
As the embedded pc-scale clumpy structure is erased, the gas reassembles into a larger warm cloud.

\begin{figure*}
	\centering
    \includegraphics[bb=100 00 1100 1070,width=15cm,scale=0.2]{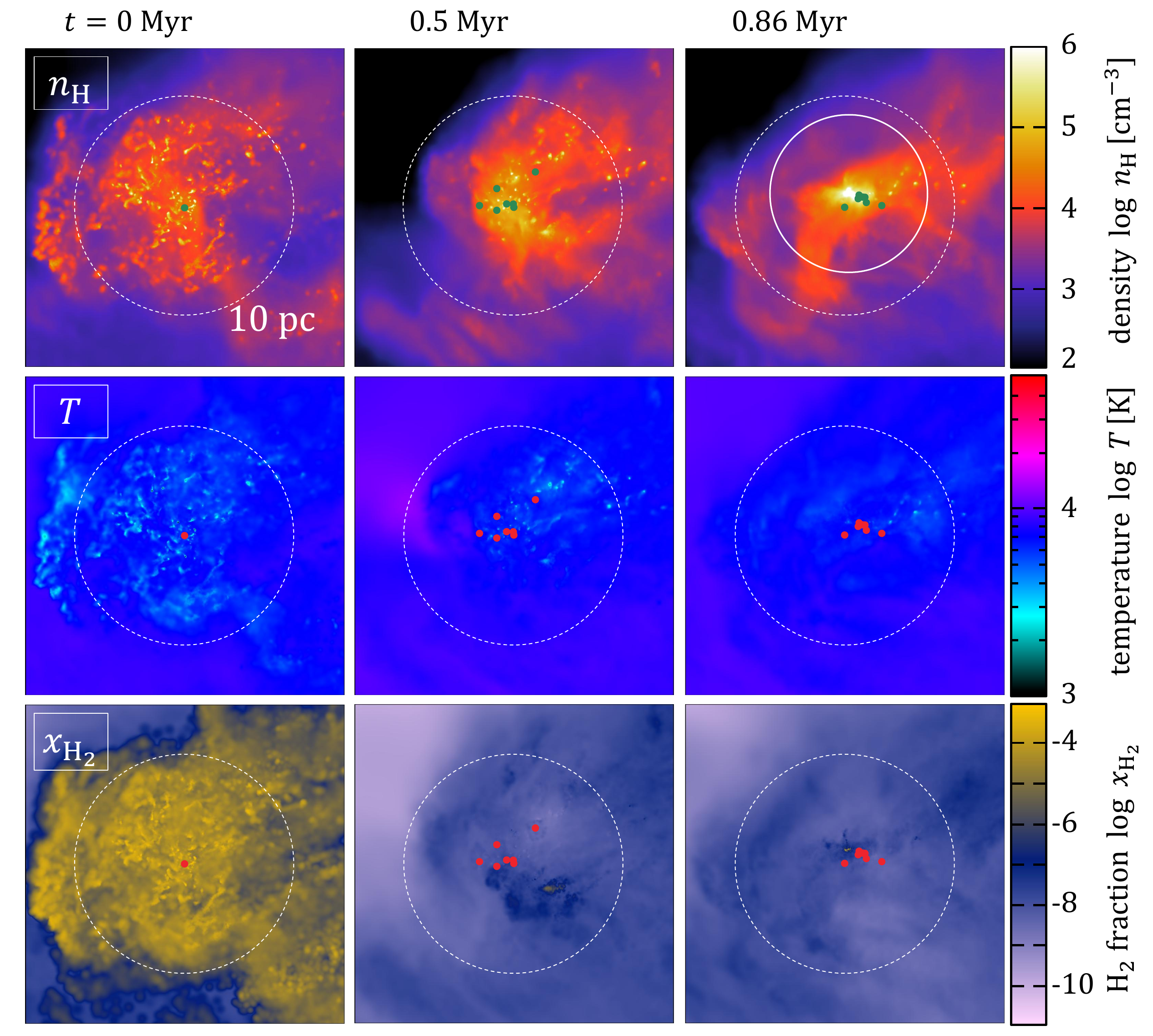}
\caption{
Mass-weighted projected maps of gas properties in the $J_{21}=10^2$ run.
From left to right, the panels correspond to $t=0$, $0.5$, and $0.86~{\rm Myr}$ after the onset of the initial star formation in the halo.
Each panel is centred on the first-formed star particle.
The three rows show density (top), temperature (middle), and H$_2$ fraction (bottom).
Green or red points indicate star particles.
Dashed white circles mark a radius of $10~{\rm pc}$ from the centre.
In the top-right panel, the white solid circle indicates the region undergoing global collapse.
Its radius is $7.2~{\rm pc}$, corresponding to the collapse radius estimated from the minimum in the $r$--$\dot{M}$ diagram
(Figure~\ref{fig:r-nH-T-Mdot}), and broadly comparable to the Jeans length of the warm gas cloud.
}
\label{fig:rho-T_splash_J1e2}
\end{figure*}

\begin{figure*}
	\centering
    \includegraphics[bb=100 00 1100 1070,width=15cm,scale=0.2]{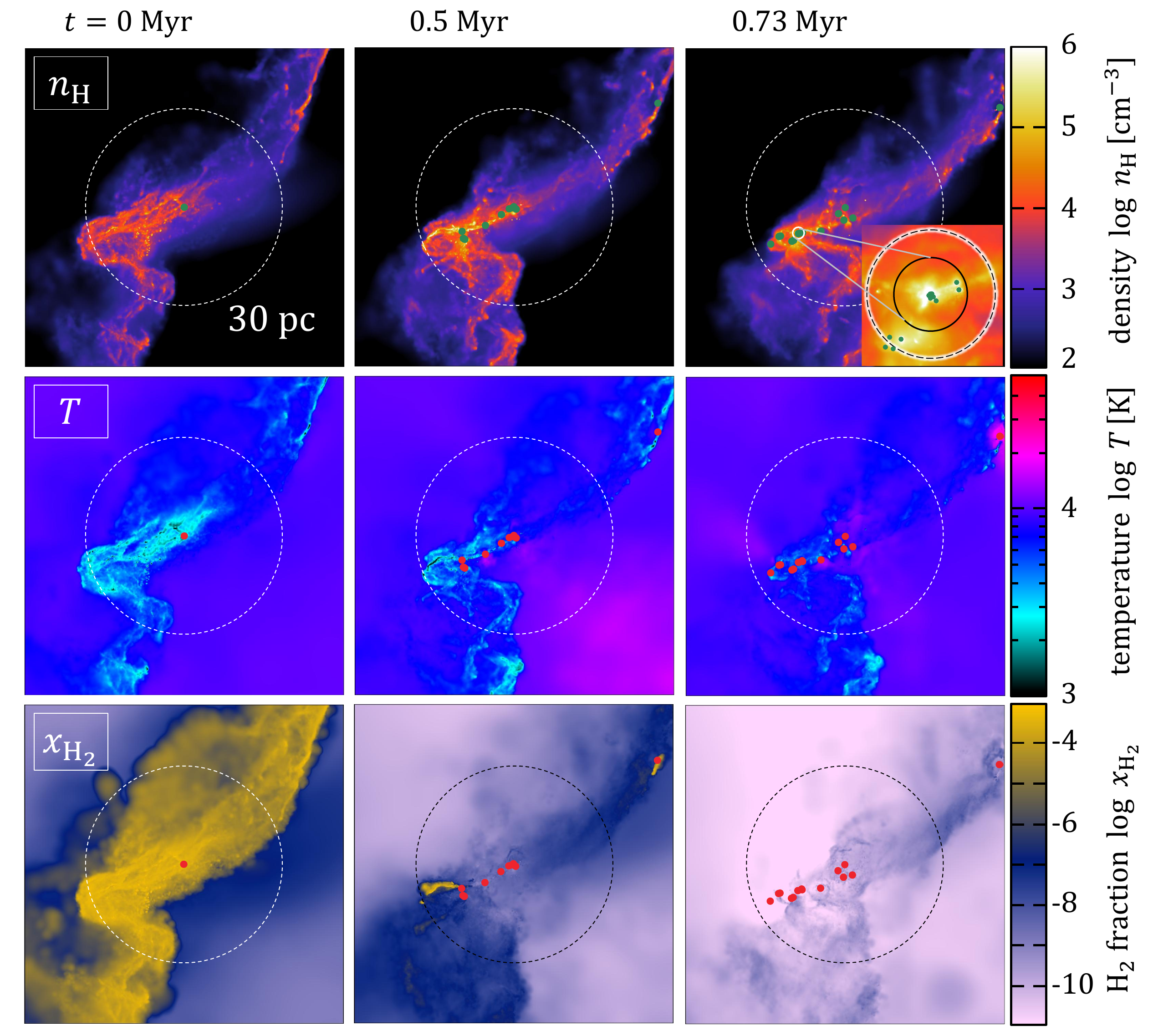}
\caption{
Same as Figure~\ref{fig:rho-T_splash_J1e2}, but for the $J_{21}=10$ run.
From left to right, the panels correspond to $t=0$, $0.5$, and $0.73~{\rm Myr}$ after the onset of star formation.
The inset in the top-right panel shows a zoomed-in view of the region undergoing global collapse, which is marked by the solid black circle with a radius of $1.7~{\rm pc}$.
}
\label{fig:rho-T_splash_J1e1}
\end{figure*}

\begin{figure*}
	\centering
    \includegraphics[bb=100 00 1100 1070,width=15cm,scale=0.2]{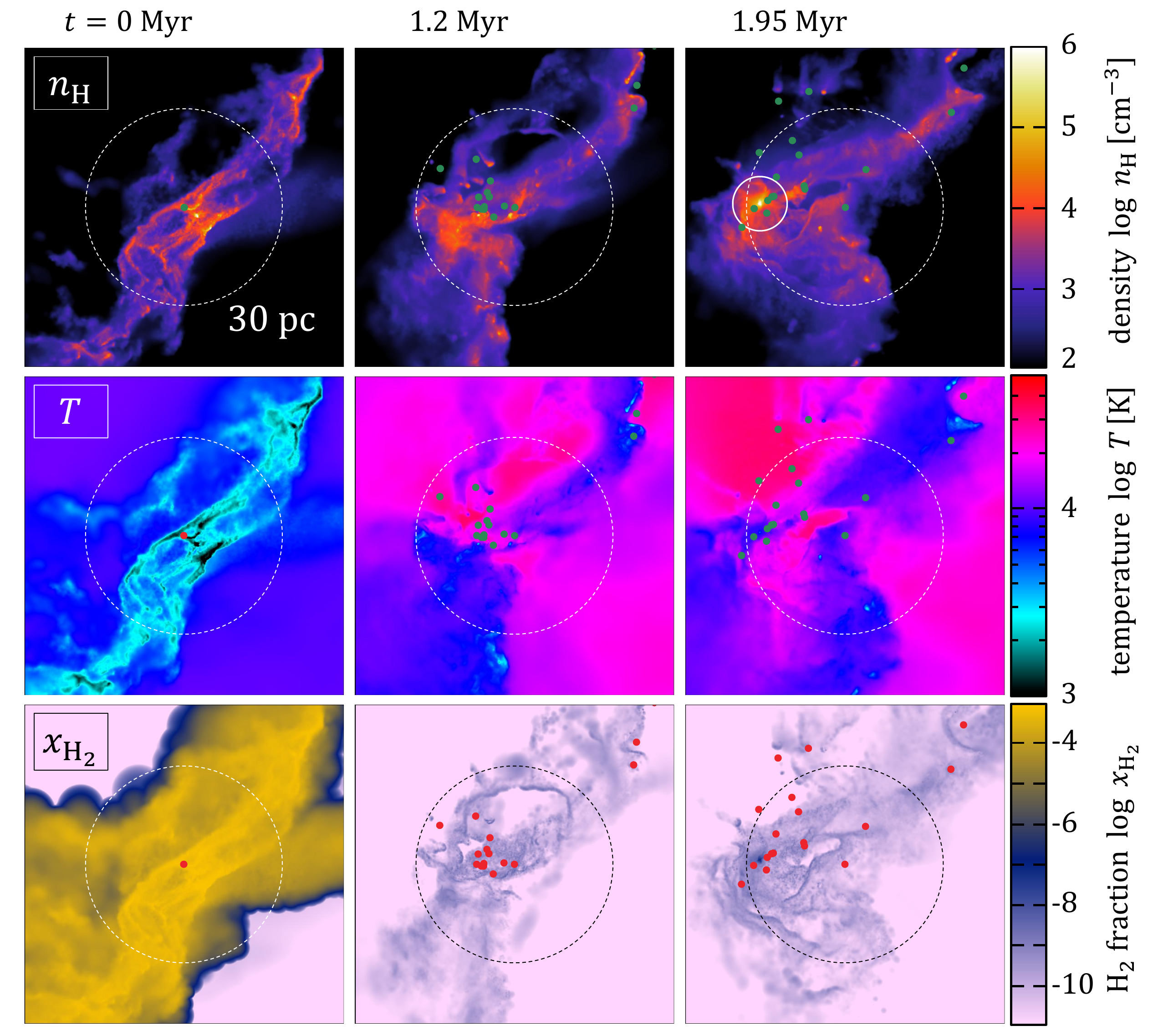}
\caption{
Same as Figure~\ref{fig:rho-T_splash_J1e2}, but for the $J_{21}=1$ run.
From left to right, the panels correspond to $t=0$, $1.2$, and $1.95~{\rm Myr}$ after the onset of star formation.
Dashed white or black circles mark a radius of $30~{\rm pc}$ from the centre.
The solid white circle mark the region that undergoes global collapse, with a radius of $8.2~{\rm pc}$.
}    
\label{fig:rho-T_splash}
\end{figure*}

The gas that has cooled and condensed through H$_2$ cooling is subsequently irradiated by the newly formed stars.
As H$_2$ is photodissociated, further molecular cooling is suppressed, and compressional heating raises the gas temperature to $T\sim8\times10^3~{\rm K}$.
At $t=0.5~{\rm Myr}$ in Figure~\ref{fig:rho-T_splash_J1e2}, the gas has reassembled into a larger dense structure with $n_{\rm H}\sim10^4~{\rm cm^{-3}}$, extending over $\sim5$--$10~{\rm pc}$.
This scale is broadly consistent with the Jeans length, $\lambda_{\rm J}\sim5~{\rm pc}$, for gas with $n_{\rm H}\sim10^4~{\rm cm^{-3}}$ and $T\sim8\times10^3~{\rm K}$.
By suppressing H$_2$ cooling and thereby raising the gas temperature, internal radiative feedback thus increases the characteristic Jeans scale from $\sim1~{\rm pc}$ to $\sim5~{\rm pc}$, suppressing further small-scale fragmentation.
FUV feedback from Pop II stars in low-metallicity galaxies has similarly been found to suppress H$_2$ cooling and small-scale fragmentation, promoting the formation of a more massive warm gas structure \citep{Sugimura+2024}.

We next estimate the FUV intensity produced by the newly formed stars to assess whether it is sufficient to suppress H$_2$ cooling in the central region.
For a total stellar mass of $\sim10^3~{\rm M_\odot}$ during the early phase (Figures~\ref{fig:t-Mtot} and \ref{fig:t-Msinks}), the total UV luminosity is estimated to be $L_{\rm UV}\simeq5\times10^{40}~{\rm erg~s^{-1}}$, assuming massive Pop III stars with $T_{\rm eff}\simeq10^5~{\rm K}$ \citep{Schaerer2002}.
The corresponding local LW intensity at a distance $r$ is
\begin{eqnarray}
J_{21}
&\sim&
10^5
\left(\frac{r}{10~{\rm pc}}\right)^{-2}.
\end{eqnarray}
This exceeds the critical intensity required to suppress H$_2$ cooling for a hard Pop III spectrum,
$J_{\rm crit}(T_{\rm eff}=10^5~{\rm K})\sim10^{4-5}$
\citep{Shang+2010,Sugimura+2014}.
The resulting internal radiation field can therefore suppress H$_2$ cooling, and hence the associated fragmentation, throughout the central $\sim10~{\rm pc}$ region.

We note that the critical intensity $J_{\rm crit}$ is conventionally defined for externally irradiated collapsing clouds, using either one-zone models or three-dimensional simulations with an imposed radiation field.
The present case differs in that the radiation sources form within the cloud, and the irradiation varies in both space and time after fragmentation has already begun.
Thus, $J_{\rm crit}$ should not be regarded as a sharp quantitative threshold for the present system.
Nevertheless, the estimated stellar FUV intensity exceeds the conventional $J_{\rm crit}$, consistent with the strong suppression of H$_2$ cooling and further fragmentation seen in our simulations.

%------------------------------------------------------------------------%
%%%%% FIG.5 %%%%%
%------------------------------------------------------------------------%

%------------------------------------------------------------------------%
%%%%% FIG.6 %%%%%
%------------------------------------------------------------------------%

%------------------------------------------------------------------------%
%%%%% FIG.7 %%%%%
%------------------------------------------------------------------------%

\subsection{Transition to Atomic Cooling and Global Collapse}
\label{sec:global_collapse_transition}

From $t=0.5~{\rm Myr}$ to $0.86~{\rm Myr}$ in Figure~\ref{fig:rho-T_splash_J1e2}, the pc-scale clumpy structure
progressively disappears, while the gas is reassembled into a single, larger cloud on a scale of several parsecs.
This reassembled cloud consists predominantly of warm gas in which H$_2$ has been photodissociated by radiation from stars formed within the cloud, and continues to contract under atomic cooling and self-gravity.

\begin{figure*}
	\centering
    \includegraphics[bb=80 0 1180 850,width=15cm,scale=0.2]{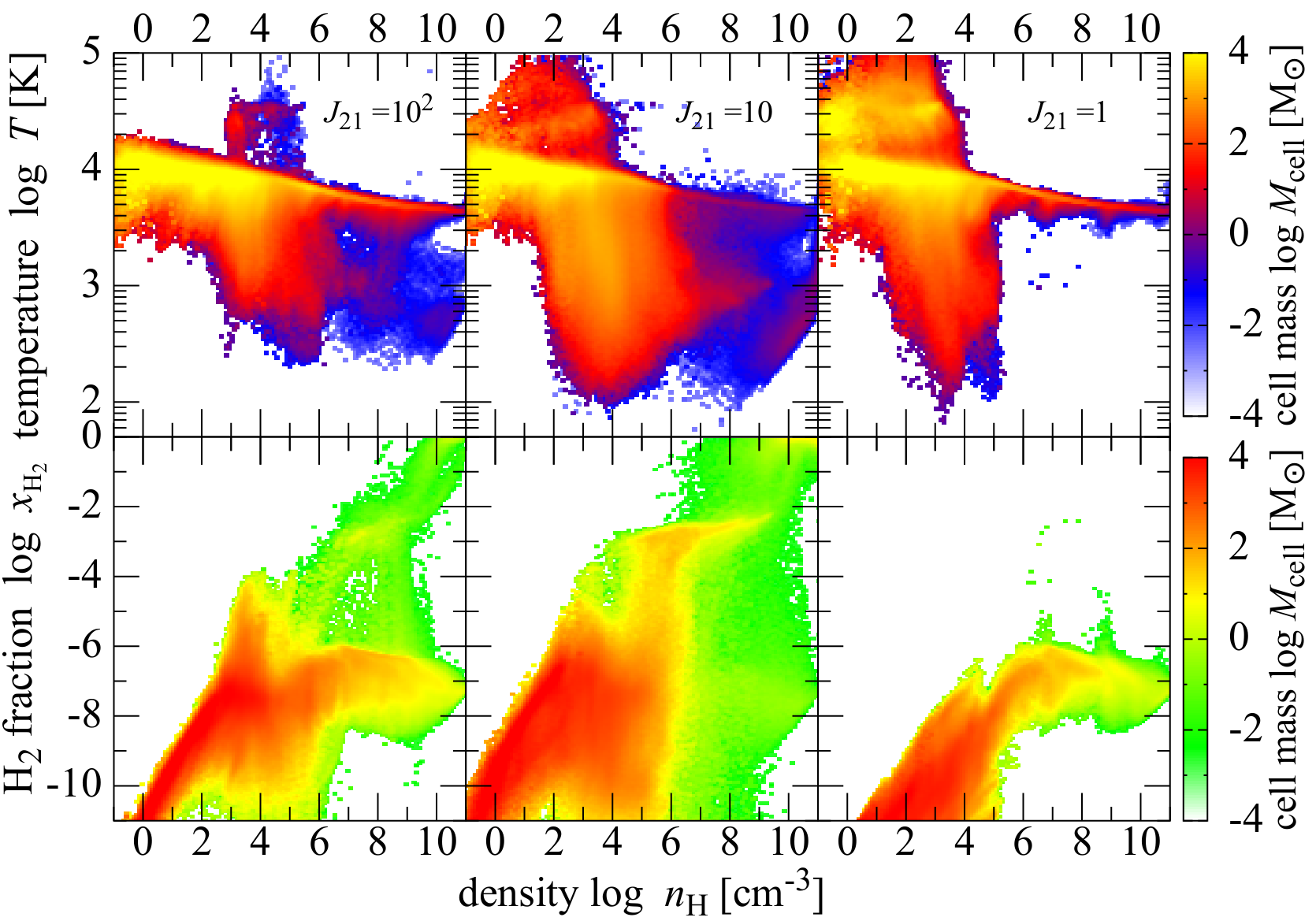}%bb=80 0 950 740,
\caption{
Gas mass distributions on the density--temperature (top) and
density--H$_2$ fraction (bottom) planes at the onset of 
star formation induced by the global collapse for the different runs.
From left to right, the panels correspond to the $J_{21}=10^2$, $10$, and $1$ runs at $t=0.86$, $0.73$, and $1.95~{\rm Myr}$, respectively.
The horizontal axis shows the gas number density, $n_{\rm H}$, while the vertical axis shows the temperature (top) or H$_2$ fraction (bottom).
The colour scale indicates the gas mass contained in each bin.
}
\label{fig:rho-T_map}
\end{figure*}

Figure~\ref{fig:rho-T_map} shows the temperature and H$_2$ fraction as functions of density during this phase for the different external FUV intensities.
In contrast to the initial-collapse phase shown in Figure~\ref{fig:rho-T_first}, where dense cores are cooled to a few hundred K by H$_2$ cooling, most of the dense gas with
$n_{\rm H}\gtrsim10^6~{\rm cm^{-3}}$ now lies along the atomic-cooling track at $T\sim8000~{\rm K}$.
The H$_2$ fraction remains at only $\sim10^{-7}$, well below the
characteristic value $x_{\rm H_2}\sim10^{-4}$ above which H$_2$ cooling becomes important \citep{YoshidaAbel2003}.
Thus, as a consequence of the suppression of H$_2$ cooling by internal radiative feedback, the global collapse proceeds along the atomic-cooling track rather than the molecular-cooling track.

The global collapse therefore represents a different mode of collapse from that during the initial-collapse phase.
During the initial collapse, the collapsing gas fragments through H$_2$ cooling into numerous dense cores.
During the global collapse, by contrast, the larger-scale warm cloud collapses as a whole without undergoing such small-scale fragmentation.
The interval between the onset of the initial star formation and the global collapse is $t\sim0.8~{\rm Myr}$.
This is comparable to the free-fall time, $t_{\rm ff}\sim1~{\rm Myr}$, of the larger-scale gas structure at $n_{\rm H}\sim10^{3-4}~{\rm cm^{-3}}$, from which the dense cores initially fragment.

All three runs eventually undergo global collapse, although the pathway leading to it varies with the external FUV intensity.
In the $J_{21}=10^2$ run discussed above, the pc-scale fragmented structure is largely erased by internal radiative feedback, while the larger-scale gas structure remains near the halo centre and is reassembled into a warm cloud.
Continued gas accumulation then drives this cloud into global collapse predominantly through atomic cooling.

The $J_{21}=10$ run follows a broadly similar evolution, but the suppression of H$_2$ cooling and fragmentation is less complete.
A large-scale self-gravitating gas structure is already present at the onset of star formation and remains near the halo centre as gas continues to accumulate.
Unlike in the $J_{21}=10^2$ run, however, some of the dense clumps produced during the initial collapse survive until the global collapse.
Figure~\ref{fig:rho-T_splash_J1e1} shows several such dense cores in the vicinity of the collapse centre.
Figure~\ref{fig:rho-T_map} also shows that some dense gas at
$n_{\rm H}\gtrsim10^4~{\rm cm^{-3}}$ retains an H$_2$ fraction of
$x_{\rm H_2}\sim10^{-3}$, allowing H$_2$ cooling to remain effective locally.
Accordingly, the $n_{\rm H}$--$T$ distribution contains a molecular-cooling component in addition to the dominant atomic-cooling track.
The large-scale gas structure nevertheless undergoes global collapse predominantly through atomic cooling, while H$_2$-cooled clumps remain locally within the cloud and continue to form stars.
Despite this persistent local fragmentation, the most massive star
ultimately grows to $m_\star>10^4~{\rm M_\odot}$.
The large-scale gas inflow can therefore feed a small number of rapidly growing stars even in the presence of local fragmentation, reminiscent of the supercompetitive accretion process discussed by \citet{ChonOmukai2020}.

In the $J_{21}=1$ run, the pathway to global collapse differs more substantially.
Radiative feedback from a massive star formed during the initial collapse drives the expansion of a large ionized bubble (Figure~\ref{fig:rho-T_splash}).
Gas near the halo centre is swept outwards and accumulates at a distance of $\sim30~{\rm pc}$ ($\sim0.03\,r_{\rm vir}$).
The accumulated gas forms a neutral cloud of size $\sim10~{\rm pc}$ near the edge of the ionized region, which subsequently undergoes global collapse.
Thus, unlike in the higher-$J_{21}$ runs, global collapse occurs in a newly assembled cloud offset from the halo centre.

Although the external FUV intensity systematically affects the overall evolution, the detailed fragmentation pattern and locations of individual star-forming clumps should not be interpreted as deterministic consequences
of $J_{21}$.
These small-scale features also depend on the stochastic evolution of fragmentation and the geometry of stellar radiative feedback.

\begin{figure*}
	\centering
    \includegraphics[bb=10 0 480 510,width=15cm,scale=0.2]{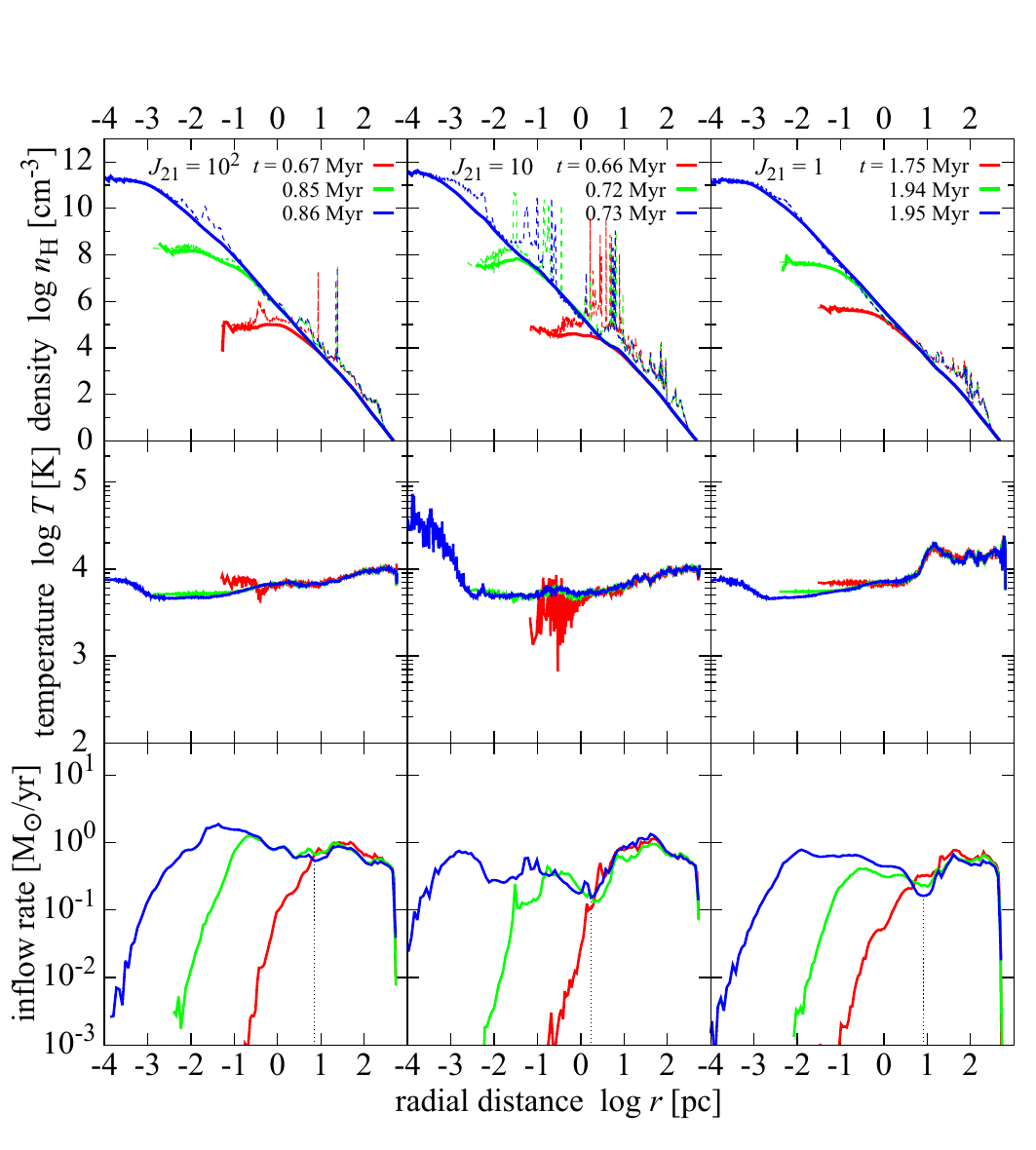}%bb=10 0 480 540
\caption{
Time evolution of the radial structure of the collapsing gas for the different runs.
From left to right, the panels correspond to the $J_{21}=10^2$, $10$, and $1$ runs.
The horizontal axis shows the radial distance from the density peak.
Red, green, and blue curves represent progressively later epochs, 
with the blue curves corresponding to the epoch immediately before star formation associated with the global collapse.
\textit{Top row:} Gas density.
Dashed curves show the mass-weighted mean density within each radial bin, while solid curves show the enclosed mean density, $\bar{n}_{\rm H}=3M_{\rm gas}(<r)/(4\pi r^3\mu\mH)$.
\textit{Middle row:} Gas temperature, shown as the mass-weighted mean within each radial bin.
\textit{Bottom row:} Gas mass inflow rate.
Dashed vertical lines mark the radii of the local minima in the inflow-rate profiles at this epoch; the regions inside these radii are defined as the global-collapse regions (see text).
}
\label{fig:r-nH-T-Mdot}
\end{figure*}

To examine how the global collapse proceeds, Figure~\ref{fig:r-nH-T-Mdot} shows the radial profiles of gas density, temperature, and mass inflow rate at three epochs for each external FUV intensity.
We first consider the density profiles.
The dashed curves show the mass-weighted mean density within each radial bin,
whereas the solid curves show the mean density enclosed within each radius,
\begin{eqnarray}
\bar{n}_{\rm H}
=
\frac{3M_{\rm gas}(<r)}
{4\pi r^3\mu m_{\rm H}},
\end{eqnarray}
where $\mu=1.2$ is the mean molecular weight and $m_{\rm H}$ is the
hydrogen mass.

The mass-weighted density profiles are sensitive to local dense clumps and therefore retain a strong signature of the fragmented structure.
Several density peaks are present around $r\sim10~{\rm pc}$, corresponding to dense clumps with $n_{\rm H}\sim10^4~{\rm cm^{-3}}$.
This density is consistent with the characteristic density at which H$_2$ cooling induces the rapid temperature decline and fragmentation during the initial collapse, as discussed in Section~\ref{sec:initial_collapse}.
By contrast, the enclosed mean density profiles average over the local clumpy structure and trace the underlying large-scale gas distribution.

At each epoch, the enclosed mean density profile consists of an approximately uniform-density central core surrounded by an envelope with $n_{\rm H}\propto r^{-2}$.
As the collapse proceeds, the central core contracts to progressively smaller radii, while the $r^{-2}$ envelope extends further inwards.
This evolution is characteristic of the self-similar runaway collapse of an isothermal gas \citep{Larson1969,Penston1969}, in which the characteristic size of the central core is approximately given by the Jeans length corresponding to the core density and temperature at each epoch.

The temperature profiles show that the collapse remains nearly isothermal at $T\simeq5$--$7\times10^3~{\rm K}$, as regulated by atomic cooling, despite an increase in density of approximately ten orders of magnitude.
The mass inflow rate rises steeply with radius within the central core and remains approximately constant at $\sim1~{\rm M_\odot~yr^{-1}}$ throughout the outer envelope.
This behaviour is consistent with the self-similar runaway collapse of an isothermal gas.

The characteristic inflow rate during such a collapse can be expressed as
\begin{eqnarray}
\dot{M}
&=&
\phi\frac{c_{\rm s}^3}{G}
\nonumber\\
&\simeq&
1.1~{\rm M_\odot~yr^{-1}}
\left(\frac{\phi}{10}\right)
\left(\frac{T}{5\times10^3~{\rm K}}\right)^{3/2}.
\end{eqnarray}
Here, $\phi$ is a dimensionless factor that depends on the dynamical state
of the collapsing cloud.
The Larson--Penston solution gives $\phi=46.9$, corresponding to a highly dynamical collapse, whereas calculations starting from more realistic initial conditions typically yield $\phi\sim10$ \citep{Foster1993}.
The resulting inflow rate of order $1~{\rm M_\odot~yr^{-1}}$ is therefore in good agreement with that found in the simulations.
The inflow-rate profile in the envelope also exhibits a local minimum around $r\sim7~{\rm pc}$. 
We use this minimum as an approximate measure of the fragmentation scale: inward mass transport from larger radii is partly redistributed among multiple collapsing cores rather than continuing coherently toward a single centre.
The resulting scale is consistent with the extent of the fragmented region seen directly in the two-dimensional density distribution in Figure~\ref{fig:rho-T_splash_J1e2}.

Despite substantial internal fragmentation, the global collapse therefore exhibits a clear Larson--Penston-type structure.
It is not simply an ensemble of independently collapsing local clumps, but instead represents the collapse of the larger underlying gas structure as a whole.

\begin{figure*}
	\centering
    \includegraphics[bb=40 0 460 200,width=15cm,scale=0.2]{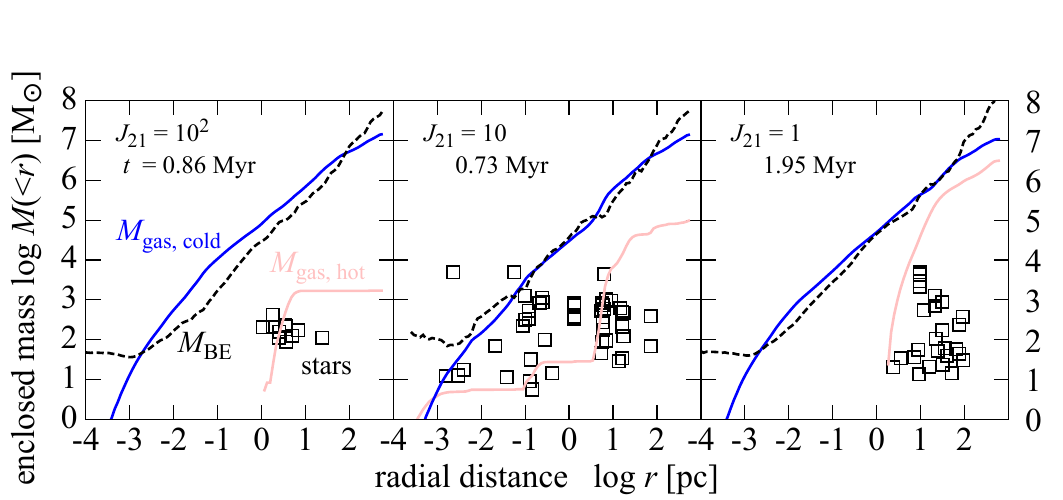}%bb=40 0 460 230,
\caption{
Radial profiles of the enclosed gas mass at the onset of star formation associated with the global collapse for the different runs.
From left to right, the panels correspond to the $J_{21}=10^2$, $10$, and $1$ runs.
The horizontal axis shows the radial distance from the density peak, and the vertical axis shows the enclosed gas mass.
The blue and pink lines show the enclosed masses of the cold
($T<2\times10^4~{\rm K}$) and hot ($T>2\times10^4~{\rm K}$) gas,
respectively.
The black dashed lines show the Bonnor--Ebert mass evaluated using the properties of the cold gas.
Square symbols indicate the positions and masses of the star particles.
}    
\label{fig:enmass}
\end{figure*}

To determine the spatial extent of the gravitationally unstable region during the global collapse, we compare the enclosed gas mass with the Bonnor--Ebert mass.
Figure~\ref{fig:enmass} shows the enclosed gas mass profiles immediately before the onset of star formation during the global collapse.
We separately evaluate the enclosed masses of the cold gas ($T<2\times10^4~{\rm K}$) and the hot, photoionized gas ($T>2\times10^4~{\rm K}$); 
(see also Figure~\ref{fig:rho-T_map}).
The mass budget is dominated by the cold gas throughout nearly the entire collapsing region.
The dashed curves show the Bonnor--Ebert mass evaluated using the properties of the cold gas, since this is the component relevant to gravitational collapse and star formation.
The hot photoionized gas is excluded from this estimate because it does not directly participate in the collapse.
Moreover, owing to the strong temperature dependence
\begin{eqnarray}
M_{\rm BE}\propto n_{\rm H}^{-1/2}T^{3/2},
\end{eqnarray}
even a small amount of hot gas can substantially raise the average temperature and lead to an overestimate of the Bonnor--Ebert mass.

The enclosed mass profiles show that, in all runs, the enclosed gas mass exceeds the Bonnor--Ebert mass over a broad radial range, indicating large-scale gravitational instability during the global collapse.
For example, in the $J_{21}=10^2$ run, this condition is satisfied over $r\sim10^{-3}$--$10^2~{\rm pc}$.
This demonstrates that the global collapse involves gravitational
instability of the large-scale gas structure, rather than only the
collapse of individual dense cores produced by fragmentation.
The gravitationally unstable region extends to $r\sim10^2~{\rm pc}$ and contains $\simeq5\times10^6~{\rm M_\odot}$ of gas, providing an approximate upper limit to the gas mass available for star formation during this phase.

The positions and masses of stars already formed by this epoch are also shown in Figure~\ref{fig:enmass}.
These stars are distributed over the radial range where small-scale H$_2$-cooled dense cores were produced by fragmentation during the initial collapse, as discussed in Section~\ref{sec:initial_collapse}.

%------------------------------------------------------------------------%
%%%%% FIG.8 %%%%%
%------------------------------------------------------------------------%

%%%%% FIG.9 %%%%%

%%%%% FIG.10 %%%%%

\subsection{Stellar Mass Distribution}
\label{sec:mass_function}

\begin{figure*}
	\centering
    \includegraphics[bb=40 0 480 320,width=15cm,scale=0.2]{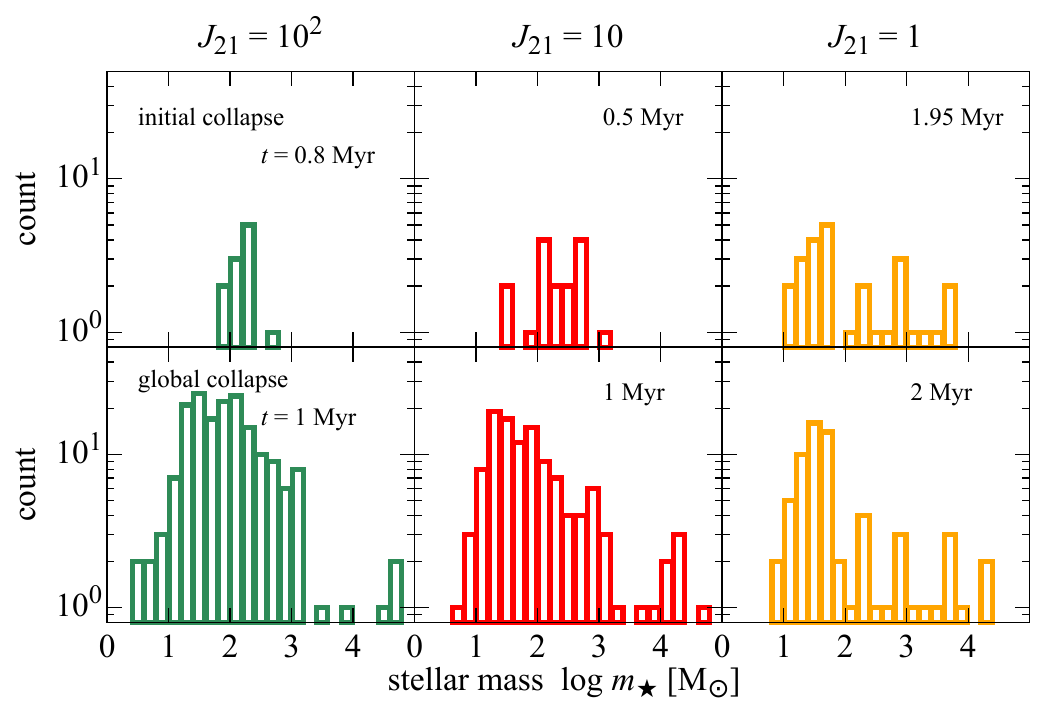}
\caption{
Stellar mass distributions for the different runs.
The horizontal axis shows the stellar mass, $m_\star$, and the vertical axis shows the number of stars in each mass bin.
From left to right, the panels correspond to the $J_{21}=10^2$, $10$, and $1$ runs.
\textit{Top row:} Mass distributions resulting from star formation during the initial-collapse phase, shown immediately before star formation associated with the global collapse.
\textit{Bottom row:} Mass distributions at the end of each simulation, following the global-collapse phase.
}
\label{fig:IMF_before}
\end{figure*}

We now examine the stellar mass distribution, with particular emphasis on the preferential growth of massive stars 
following the onset of star formation induced by the
global collapse.
Figure~\ref{fig:IMF_before} shows the stellar mass distributions 
resulting from the initial fragmentation phase (top row) and at the end of the simulations (bottom row).

We first consider the distributions produced during the initial fragmentation phase.
Immediately before star formation associated with
the global collapse, the number of stars already formed is somewhat larger for lower $J_{21}$, but remains at only $\sim10$--$20$ in all runs.

The distributions tend to extend to higher masses toward lower $J_{21}$, broadly consistent with the increase in the characteristic fragmentation mass estimated in Section~\ref{sec:initial_collapse}.
However, each mass bin contains at most only a few stars, so the detailed differences among the three histograms are subject to substantial stochasticity and are difficult to interpret as a systematic dependence on $J_{21}$.

Following the onset of star formation induced by the global collapse,
the high-mass end extends substantially in all runs, with a small number of stars growing preferentially to form SMSs with masses $>10^4~{\rm M_\odot}$.
Some of these massive stars also form gravitationally bound multiple systems.
Their possible evolution into massive BH binaries is discussed in Section~\ref{sec:transients}.
Despite the development of this high-mass tail, the peak of the mass distribution remains nearly unchanged, indicating that lower-mass star formation proceeds concurrently.
Star formation during the global collapse can therefore be characterized by the continued formation of lower-mass stars together with the preferential growth of a small subset into SMSs.

The final mass distributions shown in Figure~\ref{fig:IMF_before} still show noticeable differences among the runs.
The $J_{21}=10$ and $10^2$ runs both exhibit a main concentration around $\sim10^2~{\rm M_\odot}$ and a high-mass tail extending beyond $10^4~{\rm M_\odot}$, whereas the $J_{21}=1$ run contains fewer intermediate-mass stars with $m_\star\sim10^{2-3}~{\rm M_\odot}$.
The latter difference likely reflects the much shorter evolution time after the star formation induced by global collapse begins at $t\simeq1.95~{\rm Myr}$ in this run.
Because the number of stars in each bin is small and the apparent shape also depends on the adopted binning, however, the histograms alone are not well suited for a quantitative comparison of the mass distributions.
We therefore examine the cumulative stellar mass distributions below.

\begin{figure}
    \includegraphics[bb=10 10 470 210,width=15cm,scale=0.2]{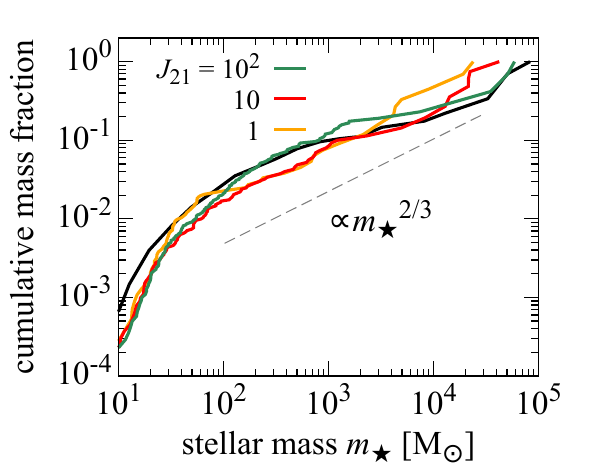}
\caption{
Cumulative stellar mass fraction at the final epoch of each run.
The horizontal axis shows the stellar mass, $m_\star$, and the vertical axis shows the cumulative stellar mass fraction, $M_{\rm cum}(<m_\star)/M_{\rm tot}$.
The green, red, and orange lines correspond to the $J_{21}=10^2$, $10$, and $1$ runs, respectively.
The black line shows the corresponding distribution for the $J_{21}=10^3$ run of \citet{ChonOmukai2025}.
The grey dashed line indicates a reference slope proportional to $m_\star^{2/3}$, with arbitrary normalization.
}
\label{fig:cumulative_fraction}
\end{figure}

We next examine how stars in different mass ranges contribute to the total stellar mass budget.
We define the cumulative stellar mass contained in stars below a given mass $m_\star$ as
\begin{eqnarray}
M_{\rm cum}(<m_\star)
\equiv
\int_{0}^{m_\star} m'_\star \frac{dN}{dm'_\star} dm'_\star.
\end{eqnarray}
Figure~\ref{fig:cumulative_fraction} shows this quantity normalized by the total stellar mass, $M_{\rm cum}(<m_\star)/M_{\rm tot}$.
The stellar mass budget is dominated by a small number of massive stars at the high-mass end.
In all runs, approximately $80$--$90~\%$ of the total stellar mass is contained in stars with $m_\star>10^3~{\rm M_\odot}$.
This behaviour is characteristic of oligarchic growth, in which the global collapse feeds a small number of massive stars.

Despite the differences visible in the binned histograms in Figure~\ref{fig:IMF_before}, the cumulative profiles at $m_\star\lesssim10^3~{\rm M_\odot}$ have remarkably similar shapes among the runs.
All profiles rise steeply below $\sim10^2~{\rm M_\odot}$, reflecting the low-mass cutoff of the stellar mass distribution, and approximately follow $M_{\rm cum}(<m_\star)\propto m_\star^{2/3}$ at higher masses.
This suggests that the differences in the stellar populations produced during the initial fragmentation phase do not leave a strong systematic imprint on the final mass distribution at $m_\star\lesssim10^3~{\rm M_\odot}$, even though they can affect the radiative-feedback history leading to the global collapse.
Once the large-scale collapse becomes dominated by atomic cooling, its subsequent evolution is governed primarily by similar thermodynamics among the runs, which may account for the similarity of their stellar
mass distributions.
The black line in Figure~\ref{fig:cumulative_fraction} extends this comparison to a stronger external FUV field, showing the cumulative
mass distribution reconstructed from the stellar mass function of the
$J_{21}=10^3$, $Z=10^{-6}~{\rm Z_\odot}$ run with radiative feedback
in \citet{ChonOmukai2025} at $t=2~{\rm Myr}$.
Its overall shape is broadly similar to those found in our simulations.
This similarity suggests that, once H$_2$ photodissociation becomes sufficiently effective to regulate the thermal evolution of the collapsing gas, the resulting stellar mass distribution may be relatively insensitive to the precise intensity of the external FUV field.

The slope of this cumulative mass function can be related to the commonly used differential IMF slope.
If the high-mass stellar mass distribution is written as
\begin{eqnarray}
\frac{dN}{dm_\star}&\propto& m_\star^{-\alpha},
\end{eqnarray}
then
\begin{eqnarray}
M_{\rm cum}(<m_\star)
&\propto&
m_\star^{2-\alpha}.
\end{eqnarray}
Thus, the relation
$M_{\rm cum}(<m_\star)\propto m_\star^{2/3}$
found in our simulations corresponds to $\alpha\simeq 4/3$.
This is substantially more top-heavy than the canonical Salpeter-like high-mass slope of $\alpha\simeq2.35$, indicating that the resulting stellar population is strongly weighted towards massive stars.

The characteristic mass scale at the low-mass end of the stellar distribution is related to the H$_2$-cooled fragmentation during the initial collapse.
As discussed in Section~\ref{sec:initial_collapse}, this scale is approximately set by the Jeans mass near the temperature minimum and increases toward weaker external FUV irradiation.
This broadly accounts for the mass range populated by stars formed during the initial star-formation episode.

The location of this peak is also close to the numerical resolution limit of our sink treatment.
For reference, the Bondi radius is
$r_{\rm Bondi}\simeq2.5\times10^{-3}~{\rm pc}
(m_\star/100~{\rm M_\odot})
(T/10^4~{\rm K})^{-1}$,
and becomes comparable to or larger than the adopted sink radius for
$m_\star\sim10$--$10^2~{\rm M_\odot}$,
depending on the gas temperature.
The detailed shape and characteristic mass at the low-mass end may therefore be affected by numerical resolution.
This uncertainty has little direct impact on the final stellar mass budget, since stars around $\sim10^2~{\rm M_\odot}$ account for only $\sim1~\%$ of the total stellar mass in Figure~\ref{fig:cumulative_fraction}.

We also note that unresolved fragmentation during the initial star-formation episode could indirectly affect the subsequent evolution by altering the strength of radiative feedback.
More efficient fragmentation into lower-mass stars would reduce the radiative output per unit stellar mass, requiring a larger total stellar mass to build up before radiative feedback becomes strong enough to suppress further fragmentation driven by H$_2$ cooling.
The resulting increase in the total stellar mass may partially compensate for the reduced radiative output per unit stellar mass.

%------------------------------------------------------------------------%
%%%%% FIG.11 %%%%%

%%%%% FIG.12 %%%%%
%------------------------------------------------------------------------%

\section{DISCUSSION}\label{sec:discussion}

The results presented above suggest a pathway to SMS formation that does not require the extremely strong external FUV irradiation assumed in the conventional direct-collapse scenario.
Once the first Pop~III stars form, their internal radiative feedback can suppress H$_2$ cooling and trigger the global collapse of the surrounding atomic-cooling gas.

In the following, we discuss the subsequent growth and fate of the most massive stars
(Section~\ref{sec:implications}), their implications for LRDs and SMBHs
(Section~\ref{sec:LRD}), the accompanying compact stellar component
(Section~\ref{sec:compact_stellar_component}), comparisons with previous
studies (Section~\ref{sec:comparison}), and possible transient signatures
(Section~\ref{sec:transients}).

% --------------------------------------------------

\subsection{Subsequent growth and fate of the most massive stars}
\label{sec:implications}

In this subsection, we estimate the possible final masses of the most massive stars and discuss their eventual evolution into massive BHs.
The simulations end while these stars are still growing: they have reached
$m_\star\sim10^4~{\rm M_\odot}$
and continue to accrete at high rates
(see Section~\ref{sec:overview}).
Their subsequent growth and final fate therefore remain uncertain.

The high accretion rates found in our simulations suggest that the most massive stars can continue to grow
without being strongly limited by their own ionizing feedback.
Their typical accretion rates are
$\dot{m}_\star\sim0.1$--$1~{\rm M_\odot~yr^{-1}}$
(Figure~\ref{fig:t-Msinks}),
well above the critical value
$\dot{m}_{\rm crit}\simeq0.04~{\rm M_\odot~yr^{-1}}$ 
above which rapidly accreting protostars remain inflated and their UV feedback is strongly suppressed.
At such high accretion rates, 
the effective temperature stays as low as
$T_{\rm eff}\sim5000~{\rm K}$,
making ionizing feedback from the star itself weak and allowing rapid accretion to continue
\citep{Hosokawa+2012}.
If maintained for
$\sim1~{\rm Myr}$,
the accretion rates measured in our simulations would allow the stars to grow to
$\gtrsim10^5~{\rm M_\odot}$.

If rapid accretion continues, the stars remain in the SMS phase until they eventually collapse after exhausting their nuclear fuel or becoming unstable to general-relativistic effects.
Their final masses depend on the accretion rate and can be approximated as
\begin{eqnarray}
m_{\star,\rm final}
&\simeq&
\left[
0.83
\left(
\frac{\dot{m}_\star}
{1~{\rm M_\odot~yr^{-1}}}
\right)
+
2.48
\right]
\times10^5~{\rm M_\odot}
\label{eq:woods}
\end{eqnarray}
\citep{Woods+2017}.
Applying this relation to the mean accretion rates found in our simulations gives final stellar masses of
$\sim{\rm few}\times10^5~{\rm M_\odot}$.
These estimates should be regarded as optimistic, as they assume that the gas reservoir remains available and that high accretion rates are sustained throughout the lifetime of the SMSs.

The actual mass growth may terminate before these upper limits are reached because of feedback from the surrounding stellar population.
Photoionization feedback is unlikely to significantly impede the large-scale gas supply, whereas supernova feedback may become important on a timescale of
$\sim2~{\rm Myr}$.

The global-collapse region contains not only SMSs but also numerous massive stars with $m_\star\sim10^2~{\rm M_\odot}$.
Ionizing radiation from these stars heats the surrounding gas, but the high recombination rate in the dense collapsing envelope confines the ionized regions to scales of order a few pc (see Appendix~\ref{app:photoionization_feedback}).

Supernova feedback is not included in the present simulations.
Massive Pop~III stars with $m_\star\gtrsim10^{2}~{\rm M_\odot}$ have characteristic lifetimes of $\sim2~{\rm Myr}$ \citep[e.g.][]{Schaerer2002},
after which supernova explosions may affect the central gas reservoir.
Their impact cannot be determined from our simulations.
We therefore consider two limiting estimates, referred to below as the conservative and optimistic cases.

In the conservative case, we assume that supernova feedback terminates further stellar growth at $t=2~{\rm Myr}$.
We extrapolate the stellar growth to this time using the mean accretion rates inferred from Figure~\ref{fig:t-Msinks}.
If the extrapolated mass exceeds the upper limit given by equation~(\ref{eq:woods}), we adopt the latter value.

In the optimistic case, we assume that the gas reservoir survives supernova feedback and continues to supply gas to the most massive stars until
the upper mass limit given by equation~(\ref{eq:woods}) is reached.
The resulting final stellar masses are summarized in Table~\ref{tab:final_mass_estimates}.

In the conservative case, whether the most massive stars can reach the $\sim10^5~{\rm M_\odot}$ scale depends strongly on how early the system enters the SMS formation phase.
For the $J_{21}=10^2$ run, extrapolation to $t=2~{\rm Myr}$ gives $m_\star\sim5\times10^5~{\rm M_\odot}$.
Since this exceeds the upper limit from equation~(\ref{eq:woods}), we adopt the value given by that relation.
For $J_{21}=10$, the extrapolated mass reaches $\sim10^5~{\rm M_\odot}$ by $t=2~{\rm Myr}$.
In the $J_{21}=1$ run, the transition to the SMS formation phase occurs much later, at $t=1.95~{\rm Myr}$, leaving little time for subsequent growth before the assumed onset of supernova feedback.
The extrapolated mass therefore remains close to the value at the end of the simulation, $\sim2\times10^4~{\rm M_\odot}$.

Taken together, these estimates indicate that the most massive stars can reach
$m_\star\gtrsim10^5~{\rm M_\odot}$ if the gas supply persists sufficiently long.
Their final masses are determined primarily by whether supernova feedback terminates the gas supply before they reach the stellar-evolution limit given by equation~(\ref{eq:woods}).

%----------------------------------------------------------
\begin{table}
\centering
\caption{
Estimated final masses of the most massive stars.
In the conservative case, accretion is extrapolated to
$t=2~{\rm Myr}$, when supernova feedback is assumed to terminate further growth, with the final mass capped at the upper limit given by equation~(\ref{eq:woods}).
In the optimistic case, accretion is assumed to continue until the upper mass limit given by equation~(\ref{eq:woods}) is reached.
}
\label{tab:final_mass_estimates}

\begin{tabular}{lccc}
\hline
Run
&
$\langle\dot{m}_\star\rangle$
&
$m_{\rm *,final}$
(conservative)
&
$m_{\rm *,final}$
(optimistic)
\\
&
[${\rm M_\odot~yr^{-1}}$]
&
[${\rm M_\odot}$]
&
[${\rm M_\odot}$]
\\
\hline
$J_{21}=10^2$ & $0.41$ & $2.8\times10^5$ & $2.8\times10^5$ \\
$J_{21}=10$ & $8.1\times10^{-2}$ & $1.0\times10^5$ & $2.5\times10^5$ \\
$J_{21}=1$ & $0.46$ & $2.3\times10^4$ & $2.9\times10^5$\\
\hline
\end{tabular}
\end{table}
%----------------------------------------------------------

% ----------------------------------------------- %

\subsection{Implications for Little Red Dots and the origin of SMBHs}
\label{sec:LRD}

In this subsection, we discuss the implications of our SMS/DCBH formation channel for LRDs and the present-day SMBH population. 
We first compare the expected BH masses and abundance with those inferred for LRDs, and then compare the estimated DCBH abundance with the number density of present-day SMBHs.

To assess whether the DCBHs formed through our scenario can power LRDs with comparable luminosities, we first estimate their subsequent growth through gas accretion.
The SMS collapse is expected to leave a BH containing most of the stellar mass \citep[e.g.,][]{ShibataShapiro+2002}.
Figure~\ref{fig:enmass} shows that the self-gravitating gas reservoir surrounding the stellar system extends over $\sim10^2~{\rm pc}$ and contains $\sim4\times10^6~{\rm M_\odot}$.
If this reservoir is not disrupted by radiative or supernova feedback, its ongoing self-gravitating inflow may continue to feed the central BH after SMS collapse \citep{Chon+2026}.
The fraction of this reservoir ultimately accreted on to the BH is uncertain. For an order-of-magnitude estimate, we adopt an accreted fraction of $\sim30\%$, motivated by the finding that the most massive star acquires $20$--$40\%$ of the total stellar mass in our simulations.
This gives a BH mass of $\sim10^6~{\rm M_\odot}$.

This is broadly consistent with the $\sim10^6~{\rm M_\odot}$ BH mass scale inferred for some LRDs if their
emission is powered by accreting BHs
\citep{Naidu+2025,Kido+2025,InayoshiMaiolino2025,  BegelmanDexter2026,deGraaff+2025}.
However, substantially more massive BHs have also been found in LRDs. In particular, a recent dynamical measurement for A2744-QSO1 indicates $M_{\rm BH}\sim5\times10^7~{\rm M_\odot}$ \citep{Juodzbalis+2026}.

We next estimate the cosmological abundance of DCBHs formed through this channel and compare it with the observed abundances of LRDs and present-day SMBHs.

For SMS/DCBH formation through this channel, the gas must remain pristine until the halo reaches the atomic-cooling scale,
$M_{\rm halo}\sim10^8~{\rm M_\odot}$.
For this estimate, we classify a halo as non-pristine if star formation occurs in any of its progenitor minihaloes.
At external FUV intensities comparable to the cosmological mean background, $J_{21}\sim6$--$40$ \citep{Dijkstra+2008},
the merger-tree calculations of \citet{Fernandez+2014} give pristine fractions of
\begin{eqnarray}
f_{\rm pristine}
&\sim&
2\times10^{-4},\quad 4\times10^{-3},
\end{eqnarray}
for $J_{\rm 21}=10$ and $30$, 
respectively.

The abundance of atomic-cooling haloes with $M_{\rm halo}\sim10^8~{\rm M_\odot}$ at $z\lesssim10$ is approximately
\begin{eqnarray}
n_{\rm ACH}
&\sim&
10~{\rm cMpc^{-3}}
\end{eqnarray}
\citep[e.g.][]{BarkanaLoeb2001}.
Combining this with the pristine fractions estimated above,
we obtain
\begin{eqnarray}
n_{\rm DCBH}
\sim f_{\rm pristine}n_{\rm ACH}
&\sim&
2\times10^{-3},
4\times10^{-2}~{\rm cMpc^{-3}}
\end{eqnarray}
for $J_{\rm 21}=10$ and $30$, respectively.
Thus, even when restricting SMS/DCBH formation to strictly metal-free gas, 
the expected DCBH number density is $\sim10^{-3}$--a few $\times10^{-2}~{\rm cMpc^{-3}}$
under a cosmological-mean FUV background.

We next compare the estimated DCBH abundance with the observed LRD population.
LRDs have a number density of
$n_{\rm LRD}\sim10^{-5}$--$10^{-4}~{\rm cMpc^{-3}}$
\citep{Matthee+2024,Perez-Gonzalez+2024,Kokorev+2024,Greene+2024},
which is well below our estimated DCBH abundance.
If only a fraction
$f_{\rm duty}$
of DCBHs are observed in an LRD-like active state at any given time, their number densities are related by
\begin{eqnarray}
n_{\rm LRD}
&\sim&
f_{\rm duty}n_{\rm DCBH}.
\end{eqnarray}
Here, $f_{\rm duty}$ denotes the duty cycle of the LRD-like phase.
As in AGN populations more generally, duty cycles are naturally expected to be below unity \citep{Schawinski+2015,Aird+2018}.
The observed LRD abundance and estimated DCBH abundances then imply a characteristic duty cycle of $f_{\rm duty}\sim10^{-2}$.
If the LRD state corresponds to a single evolutionary phase in the growth of each DCBH, this duty cycle corresponds to a characteristic duration of
\begin{eqnarray}
t_{\rm LRD}
&\sim&
f_{\rm duty}t_{\rm H}
\sim10~{\rm Myr},
\end{eqnarray}
where
$t_{\rm H}\sim1~{\rm Gyr}$
is the Hubble time at
$z\sim6$.
A duration of $\sim10~{\rm Myr}$ is consistent with the picture of LRDs as a transient phase of BH-powered emission during BH growth
\citep{Naidu+2025,Kido+2025,BegelmanDexter2026}.

The inferred duty cycle also has implications for the DCBH occupation fraction of the galaxy population.
Comparison between the UV luminosity functions of LRDs and galaxies at $z\sim5$--$6$ gives
\begin{eqnarray}
n_{\rm LRD}
&\sim&
f_{\rm duty}f_{\rm occ}n_{\rm gal},
\end{eqnarray}
with
$f_{\rm duty}f_{\rm occ}\sim10^{-2}$
\citep{Asada+2026},
where $f_{\rm occ}$ is the DCBH occupation fraction.
If the same duty cycle inferred above applies, $f_{\rm duty}\sim10^{-2}$
implies an occupation fraction of order unity,
$f_{\rm occ}\sim1$.
This would suggest that most galaxies in the luminosity range relevant to LRDs already host heavy seed BHs, while only a small fraction appear as LRDs at any given time.

The inferred DCBH abundance is also comparable to that of SMBHs in the local Universe,
$n_{\rm SMBH}\sim10^{-3}$--$10^{-2}~{\rm cMpc^{-3}}$
\citep{GreeneHo2007,SchulzeWisotzki2010,Miller+2015,Zou+2025}.
This suggests that our formation channel can provide a sufficiently abundant progenitor population for present-day SMBHs.
Whether individual DCBHs subsequently grow into SMBHs is not addressed here and depends on 
their
gas accretion \citep{Latif+2018,Chon+2021,Chon+2026} 
and merger histories \citep{Valiante+2016,Trinca+2022,Schneider+2023}.
Moreover, the pristine fraction depends on the halo mass and redshift, so the DCBH abundance estimated above should be regarded as approximate.

The abundance estimated above may represent a lower limit because DCBH formation is not necessarily restricted to strictly metal-free gas.
Under strong FUV irradiation, central objects can grow into SMSs even at metallicities of $Z\lesssim10^{-3}~Z_\odot$, despite fragmentation induced by dust cooling \citep{ChonOmukai2020,ChonOmukai2025}.
Cosmological semi-analytic calculations incorporating the super-competitive accretion channel at finite metallicity likewise find an enhanced population of massive BH seeds \citep{Sassano+2021}.
However, it remains unclear whether the halo-internal irradiation mechanism studied here remains effective under such conditions.

As the metallicity increases, metal cooling causes the gas to depart from the atomic-cooling temperature of several thousand K at progressively lower densities. 
Fragmentation then occurs at lower densities and on corresondingly larger spatial and mass scales, where the free-fall time is longer.
This may delay the reassembly and subsequent global collapse of the gas, allowing stellar feedback to disrupt the cloud before global collapse is completed.
Radiation-hydrodynamic simulations including finite metallicity are required to determine whether this formation channel remains viable.

The strong increase in the number of potential formation sites at finite metallicity is also suggested by \citet{Chiaki+2023}.
Combining cosmological N-body simulations with a semi-analytic galaxy model, they estimated the abundance of SMS/DCBH formation sites while following the metallicity and internal and external LW radiation fields
of individual haloes.
They assumed DCBH formation in atomic-cooling haloes
($T_{\rm vir}>8000~{\rm K}$) with $Z<Z_{\rm th}$ and an internal radiation field of $J_{21}>10^3$.
The resulting DCBH number density increases from
$n_{\rm DCBH}\simeq0.03$--$0.06~{\rm Mpc^{-3}}$ for
$Z_{\rm th}\le10^{-4}Z_\odot$ to
$n_{\rm DCBH}\simeq1~{\rm Mpc^{-3}}$ for
$Z_{\rm th}=10^{-3}Z_\odot$.
Although their assumed radiation threshold is much higher than the external FUV intensities considered here, their results illustrate how allowing finite metallicity can substantially increase the abundance
of potential SMS/DCBH formation sites.

%%%%% FIG.13 %%%%%
\begin{figure}
    \includegraphics[bb=10 25 600 270,width=15cm,scale=0.2]{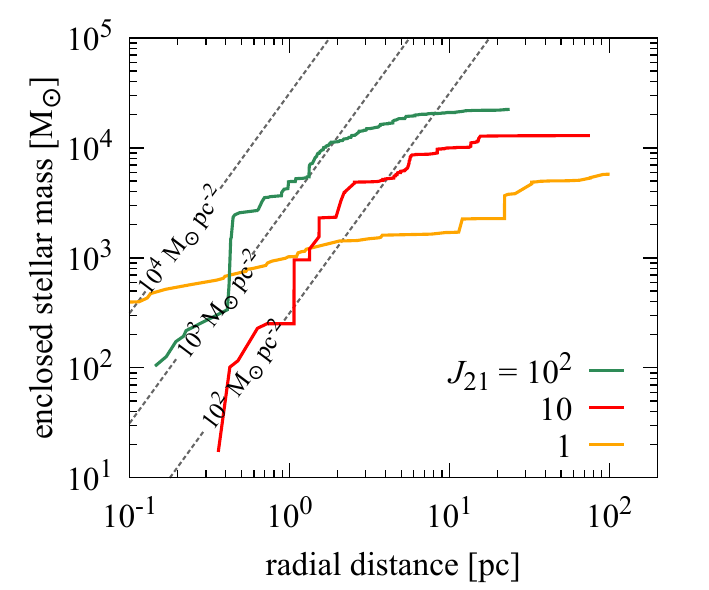}
\caption{
Radial distribution of stars in the final snapshot of each run.
The horizontal axis shows the radial distance from the most massive star, and the vertical axis shows the enclosed stellar mass, $M_{\star,\rm enc}$, including only stars with $M_\star<10^3~{\rm M_\odot}$.
The green, red, and orange solid lines correspond to the
$J_{21}=10^2$, $10$, and $1$ runs, respectively.
The grey dashed lines indicate constant stellar surface densities, $\Sigma_\star=M_{\star,\rm enc}/(\pi r^2)$, of $10^2$, $10^3$, and $10^4~{\rm M_\odot~pc^{-2}}$.
}
\label{fig:stellar_enclosed_mass}
\end{figure}

%------------------------------------------------------------------------%
\subsection{Compact Stellar Systems and Pop III Galaxy Candidates}
\label{sec:compact_stellar_component}

In this subsection, we compare the compact stellar component formed in our simulations with observed compact stellar systems and candidate Pop~III or extremely metal-poor stellar populations at high and intermediate redshifts.
We first characterize its spatial extent and stellar surface density directly from the simulations.
We then focus on the young massive-star component with
$m_\star<10^3~{\rm M_\odot}$
formed around the SMSs and use simple SED models to estimate the stellar mass required to reproduce the observed UV continuum fluxes and recombination-line luminosities.

To compare the simulated stellar systems with the compact sources observed by JWST, we first determine their characteristic size and the mass of the compact young stellar component that dominates their radiative output.
Figure~\ref{fig:stellar_enclosed_mass} shows the radial distribution of stars with $m_\star<10^3~{\rm M_\odot}$ in the final snapshot of each run, centred on the most massive star.
We define the mass of this component as
\begin{eqnarray}
M_{\star,{\rm comp}}
\equiv
M_{\star,{\rm tot}}(m_\star<10^3~{\rm M_\odot}).
\end{eqnarray}
At the end of the simulations,
$M_{\star,{\rm comp}}\sim10^4~{\rm M_\odot}$.

This mass may increase if star formation continues beyond the simulated period.
The total stellar mass growth rate after the global collapse is $\sim0.1$--$1~{\rm M_\odot~yr^{-1}}$, so sustained star formation for a few Myr could increase the total stellar mass to $\sim10^6~{\rm M_\odot}$.
Assuming that $\sim10\%$
of this mass remains in stars with
$m_\star<10^3~{\rm M_\odot}$
then gives
$M_{\star,{\rm comp}}\sim10^5~{\rm M_\odot}$.

Figure~\ref{fig:stellar_enclosed_mass} shows that the stellar component is compact, extending over $\sim1$--$10~{\rm pc}$, with most stars concentrated within the central $\lesssim10~{\rm pc}$.
In the $J_{21}=10^2$ run, the stellar surface density reaches $\Sigma_\star\gtrsim10^3~{\rm M_\odot\,pc^{-2}}$ within $\lesssim1~{\rm pc}$.
In the $J_{21}=10$ run, $\Sigma_\star\sim10^2~{\rm M_\odot\,pc^{-2}}$ within $\lesssim3~{\rm pc}$.
In the $J_{21}=1$ run, the simulation is terminated shortly after the onset of star formation induced by the global collapse.
Stars with a total mass of $\sim400~{\rm M_\odot}$ are already concentrated near the collapse centre, reaching a local stellar surface density of $\sim10^4~{\rm M_\odot\,pc^{-2}}$.
Thus, although the stellar components remain modest in total mass at the end of the simulations, they already form compact stellar systems on pc scales.

We compare the UV continuum and recombination-line luminosities, which are more directly observable than the total stellar mass and less sensitive to the assumed IMF.

We first consider the UV continuum from the compact young stellar component.
Based on the metal-free ZAMS models of \citet{Schaerer2002}, we approximate the stars in this component as blackbody emitters with
$T_{\rm eff}=10^5~{\rm K}$
and
$L_\star=0.7L_{\rm Edd}$.
These values are representative of
$m_\star\sim500$--$1000~{\rm M_\odot}$
stars, which dominate the mass and radiative output of this component in our simulations.
The total bolometric luminosity is therefore
\begin{eqnarray}
L_{\rm bol}
&=&
0.7L_{\rm Edd}(M_{\star,{\rm comp}})
\nonumber\\
&\simeq&
1.1\times10^{42}
\left(\frac{M_{\star,{\rm comp}}}{10^4~{\rm M_\odot}}\right)
{\rm erg~s^{-1}}.
\end{eqnarray}

Table~\ref{tab:massive_stellar_mass_comparison} summarizes the comparison between the characteristic size and $M_{\star,{\rm comp}}$ of the compact stellar systems in our simulations and those inferred for the observed systems.
For each observed object, we list its size and the stellar mass required to reproduce either the UV continuum or recombination-line luminosity.

Overall, the compact stellar systems formed in our simulations have sizes comparable to the pc-scale structures inferred for several JWST sources.
The simulated masses,
$M_{\star,{\rm comp}}\sim10^4~{\rm M_\odot}$,
are somewhat smaller than those inferred for several of 
the observed systems,
but continued growth to
$\sim{\rm few}\times10^4$--$10^5~{\rm M_\odot}$
could reproduce the UV or He~{\sc ii} luminosities of sources such as AMORE6 and \Hebe.
The stacked He~{\sc ii} luminosity of the HETDEX candidates discussed below requires a substantially larger mass,
$M_{\star,{\rm comp}}\sim2\times10^6~{\rm M_\odot}$,
in our simple model.

\subsubsection{UV continuum}

For the UV continuum comparison, the adopted stellar SED gives
\begin{eqnarray}
M_{\rm UV}
&\simeq&
-12.4
-
2.5\log_{10}
\left(\frac{M_{\star,{\rm comp}}}{10^4~{\rm M_\odot}}\right).
\end{eqnarray}

AMORE6 is a strongly lensed, extremely metal-poor galaxy at $z=5.7253$ \citep{Morishita+2025}.
It has an intrinsic effective radius of $R_{\rm e}\simeq36~{\rm pc}$ and a UV magnitude of
$M_{\rm UV}\simeq-14.5$.
Its size is somewhat larger than the $\sim1$--$10~{\rm pc}$ extent of the compact stellar systems in our simulations. 
Reproducing its UV luminosity with our toy SED model requires
$M_{\star,{\rm comp}}\sim7\times10^4~{\rm M_\odot}$,
which could be reached with continued star formation beyond the simulated
period.

The Firefly Sparkle is a strongly lensed galaxy at $z_{\rm spec}=8.296$ containing multiple compact stellar clumps \citep{Mowla+2024}.
Most individual clumps are unresolved, with lensing-corrected half-light radii of $R_{\rm eff}<4$--$7~{\rm pc}$, 
comparable to the $\sim1$--$10~{\rm pc}$ extent of the compact stellar systems in our simulations. 

For the brightest component, FF-6, reproducing the intrinsic F150W luminosity with our toy SED model requires $M_{\star,{\rm comp}}\simeq6\times10^5~{\rm M_\odot}$.
This is substantially larger than the current simulation value, $M_{\star,{\rm comp}}\sim10^4~{\rm M_\odot}$, and the $\sim10^5~{\rm M_\odot}$ expected from our extrapolation of continued star formation.

 At $z\simeq10.2$, \citet{Adamo+2024} report the discovery of five compact star-cluster candidates, the so-called Cosmic Gems.
Their lensing-corrected half-light radii are typically $R_{\rm eff}\sim1~{\rm pc}$, comparable to the compact stellar systems formed in our simulations.
Their inferred ages of $\sim9$--$36~{\rm Myr}$ are substantially longer than the lifetime of the massive stars considered in our SED model.
We therefore use the Cosmic Gems only as a comparison for pc-scale compactness, rather than for their stellar SEDs.

\subsubsection{Recombination-line emission}

We next estimate the He~{\sc ii} recombination-line luminosity from the compact stellar component.
The metal-free ZAMS models of \citet{Schaerer2002} give an He$^+$-ionizing photon production rate of $Q({\rm He^+})/m_\star\simeq2.7\times10^{47}~
{\rm s^{-1}~M_\odot^{-1}}$
for stars with
$m_\star=500$--$1000~{\rm M_\odot}$.
Adopting this as a representative value for the compact stellar component, we obtain
\begin{eqnarray}
Q({\rm He^+})
&\simeq&
2.7\times10^{51}
\left(\frac{M_{\star,{\rm comp}}}{10^4~{\rm M_\odot}}\right)
{\rm s^{-1}}.
\end{eqnarray}

Using the case-B conversion factor of \citet{Schaerer2002}, the He~{\sc ii}~$\lambda1640$ luminosity is
\begin{eqnarray}
L_{\rm HeII}
&=&
5.67\times10^{-12}
\left(1-f_{\rm esc,He^+}\right)
Q({\rm He^+})
\nonumber\\
&\simeq&
1.5\times10^{40}
\left(1-f_{\rm esc,He^+}\right)
\left(\frac{M_{\star,{\rm comp}}}{10^4~{\rm M_\odot}}\right)
{\rm erg~s^{-1}},
\label{eq:HeII_massive_component}
\end{eqnarray}
where, $f_{\rm esc,He^+}$ is the fraction of He$^+$-ionizing photons.
We adopt $f_{\rm esc,He^+}=0$ as our fiducial value, since these photons are expected to be efficiently absorbed within the compact stellar system.
For a non-zero escape fraction, the stellar mass required to reproduce a given He~{\sc ii} luminosity increases as
$(1-f_{\rm esc,He^+})^{-1}$.

\Hebe is an unresolved He~{\sc ii}~$\lambda1640$ emitter at $z\simeq10.6$ near GN-z11 and has been discussed as a Pop~III stellar-cluster candidate because of the absence of detected metal lines \citep{Rusta+2026,Maiolino+2026}.
Its observed He~{\sc ii}~$\lambda1640$ luminosity is
$L_{\rm HeII} \simeq 8.5\times10^{40}~{\rm erg~s^{-1}}$.
In our model, reproducing this luminosity requires $M_{\star,{\rm comp}}\simeq 6\times10^4 
\left(1-f_{\rm esc,He^+}\right)^{-1} ~{\rm M_\odot}$.

\citet{Rusta+2026} further decompose the He~{\sc ii} emission into two velocity components and identify the C1 as the more pristine Pop~III component.
Its He~{\sc ii} luminosity,
$L_{\rm HeII,C1}
\simeq
5.1\times10^{40}~{\rm erg~s^{-1}}$,
corresponds to
$M_{\star,{\rm comp}}
\simeq
3\times10^4
\left(1-f_{\rm esc,He^+}\right)^{-1}
~{\rm M_\odot}$
in our model.
For our fiducial $f_{\rm esc,He^+}=0$, this estimate is consistent with the Pop~III stellar mass inferred from the more detailed spectral
modelling of \citet{Rusta+2026}.

These masses are only a factor of $\sim3$--$6$ larger than the current simulation value of $M_{\star,{\rm comp}}\sim10^4~{\rm M_\odot}$.
Continued star formation after the global collapse could therefore produce He~{\sc ii} luminosities comparable to those of C1 and \Hebe as a whole.

\citet{Khanlari+2026} recently identified eight candidate very metal-poor or Pop III-like systems at $z\simeq2$ from the HETDEX Ly$\alpha$ emitter sample.
Their stacked spectrum shows strong He~{\sc ii}~$\lambda1640$ emission,
$L_{\rm HeII}\simeq3.0\times10^{42}~{\rm erg~s^{-1}}$,
with no significant N~{\sc v}, C~{\sc iv}, or O~{\sc iii}] emission.
Applying equation~(\ref{eq:HeII_massive_component}) to the stacked He~{\sc ii} luminosity gives
$M_{\star,{\rm comp}}\simeq 2\times10^6 \left(1-f_{\rm esc,He^+}\right)^{-1} ~{\rm M_\odot}.$
This mass is substantially larger than both the current stellar mass in our simulations and our optimistic extrapolation assuming continued star formation.
The spatial extent of the He~{\sc ii}-emitting component is not constrained by the HETDEX data.

LAP1-B is an extremely metal-poor subcomponent of the LAP1 system at $z\simeq6.64$ \citep{Vanzella+2023}.
Its H$\alpha$-emitting region has a characteristic size of only a few pc, comparable to the compact stellar systems in our simulations.
The observed H$\alpha$ luminosity corresponds to $M_{\star,{\rm comp}}\sim10^3~{\rm M_\odot}$ in our model, smaller than the current simulation value of
$\sim10^4~{\rm M_\odot}$.
No secure He~{\sc ii} detection is available, preventing a similar
comparison of the He$^+$-ionizing output.

%----------------------------------------------------------
\begin{table*}
\centering
\caption{
Summary of the comparison between the compact stellar components in our
simulations and observed compact metal-poor systems.
For our simulations, the first row shows the current snapshots, while
the second gives an optimistic extrapolation assuming continued star
formation for a few Myr.
For the observed systems, $M_{\star,{\rm comp}}$ is the stellar mass
required to reproduce the indicated observable using the simple models
adopted in this work, rather than the stellar mass inferred in the
original observational studies.
}
\label{tab:massive_stellar_mass_comparison}
\begin{tabular}{lccc}
\hline
Object
&
Radius
&
Observable
&
$M_{\star,{\rm comp}}$
\\
&
[pc]
&
&
[${\rm M_\odot}$]
\\
\hline
This work
&
$\sim1$--$10$
&
simulation snapshot
&
$\sim10^4$
\\
This work (continued star formation)
&
---
&
optimistic extrapolation
&
$\sim10^5$
\\
\hline
AMORE6~\citep{Morishita+2025}
&
$\sim36$
&
UV continuum
&
$\sim7\times10^4$
\\
The Firefly Sparkle FF-6~\citep{Mowla+2024}
&
$<4$--$7$
&
F150W continuum
&
$\sim6\times10^5$
\\
Cosmic Gems~\citep{Adamo+2024}
&
$\sim1$
&
---
&
---
\\
\Hebe (total)~\citep{Rusta+2026,Maiolino+2026}
&
$<400$
&
He~{\sc ii}~$\lambda1640$
&
$\sim6\times10^4$
\\
\Hebe C1~\citep{Rusta+2026}
&
$<400$
&
He~{\sc ii}~$\lambda1640$
&
$\sim3\times10^4$
\\
HETDEX He~{\sc ii} stack~\citep{Khanlari+2026}
&
---
&
He~{\sc ii}~$\lambda1640$
&
$\sim2\times10^6$
\\
LAP1-B~\citep{Vanzella+2023}
&
$<5$
&
H$\alpha$
&
$\sim10^3$
\\
\hline
\end{tabular}
\end{table*}

%----------------------------------------------------------

\subsection{Comparison with Previous Studies}
\label{sec:comparison}

Our main result is that SMS formation can occur under moderate external FUV irradiation even after H$_2$ cooling has already triggered fragmentation and Pop III star formation.
Radiative feedback from these stars subsequently suppress H$_2$ cooling in the surrounding atomic-cooling gas, leading to its global collapse on a timescale of
$\sim1$--$2~{\rm Myr}$.
Previous studies provide useful comparisons for different stages and aspects of this evolution.

\citet{Regan+2020} followed a primordial atomic-cooling halo exposed to a moderate FUV field for $\sim2~{\rm Myr}$.
They also find the formation of very massive stars, including one reaching $\sim6000~{\rm M_\odot}$, but their accretion rates decline as the local gas reservoirs are depleted.
Unlike in our simulations, the fragmented gas does not subsequently reassemble into a coherent $\sim10~{\rm pc}$ reservoir and undergo global collapse.
The origin of this difference remains unclear, although one potentially important distinction is that ionizing feedback was not dynamically coupled to the gas in their simulations.

\citet{Latif+2021} likewise found H$_2$ cooling leads to the formation of dense clumps and massive stars with $m_\star\sim10^3$--$10^4~{\rm M_\odot}$ in atomic-cooling haloes exposed to $J_{21}<J_{\rm crit}$.
Several of their simulations were terminated after the accretion rates onto the initially formed stars declined, before the subsequent evolution of the surrounding gas reservoir could be followed.
Our results show that the residual reservoir can continue to contract and eventually undergo global collapse.

\citet{Kiyuna+2024} found a closely related pathway in which SMS formation occurs under a moderate external FUV field through halo-internal radiative feedback.
The main difference is the trigger and timescale of the SMS-forming collapse.
In their simulation, SMS formation occurs after a
$\sim10$--$30~{\rm Myr}$
quiescent phase and is triggered by a later cosmological cold-accretion flow.
In our simulations, the residual atomic-cooling reservoir instead undergoes global collapse within
$\sim1$--$2~{\rm Myr}$, without requiring a later cold-accretion event to trigger the collapse.

\citet{Sullivan+2025} found that halo-internal FUV radiation suppresses H$_2$ cooling in a neighbouring $\sim10^3~{\rm M_\odot}$ clump, reducing its accretion rate and limiting the stellar mass to $\sim10^2$--$10^3~{\rm M_\odot}$.
Similar suppression of H$_2$ cooling occurs in the initially fragmented clumps in our simulations.
The SMSs in our simulations, however, form later through the global collapse of the much larger $\sim10~{\rm pc}$,
$\sim10^5$--$10^6~{\rm M_\odot}$
atomic-cooling reservoir.
Thus, while internal radiative feedback suppresses stellar growth on clump scales, it can subsequently promote SMS formation through the global collapse of the surrounding reservoir.

\subsection{Possible Transient Signatures}
\label{sec:transients}

The stellar systems formed in our simulations contain SMS progenitors together with numerous surrounding stars. After the SMSs collapse into massive BHs, these systems may produce electromagnetic and gravitational-wave transients through tidal disruption events (TDEs) and mergers of massive BH binaries \citep{FialkovLoeb2017,Kar+2024}.

The lower-mass stars formed around the SMSs, whose lifetimes exceed that of the SMS progenitor, provide potential sources of TDEs after the SMS collapses into a DCBH.
In the compact stellar systems expected around newly formed DCBHs, dynamical relaxation can scatter such stars onto loss-cone orbits on a timescale shorter than their lifetimes.
Few-body interactions may provide an additional channel for bringing stars close to the DCBH.
Such tidal disruptions can produce luminous transients, including long-duration soft X-ray emission accompanied by relativistic jets and radio afterglows \citep{KashiyamaInayoshi2016}.

The potential TDE sources in our simulations span a range of stellar
masses, although the most massive stars are unlikely to survive until after DCBH formation.
The peak fallback rate and jet luminosity scale approximately as
\begin{eqnarray}
\dot{M}_{\rm fb,peak},\,L_j
&\propto& m_\star^2R_\star^{-3/2}
\end{eqnarray}
\citep{Stone+2013}.
Surviving stars at the higher-mass end could therefore produce higher peak fallback rates and more luminous X-ray and radio transients, provided that they remain compact.
For bloated protostars, however, their larger radii would increase the fallback timescale and reduce the peak fallback rate and luminosity.

Our simulations also form multiple SMSs rather than a single SMS, so several massive BHs may remain within the same halo after stellar collapse.
Recent JWST observations have identified candidate systems containing multiple accreting BHs or closely separated LRDs within individual high-redshift galaxies \citep{Ubler+2026,Yanagisawa+2026}, suggesting that such multiple-BH configurations may occur at early cosmic times.
If the BHs formed in our scenario subsequently form binaries and merge, they would constitute a potential source of gravitational waves.
For example, in the $J_{21}=10^2$ run, the two most massive stars at the end of the simulation form a gravitationally bound pair with masses of $6\times10^4~{\rm M_\odot}$ and $5\times10^4~{\rm M_\odot}$, separated by $\sim0.04~{\rm pc}$.
Both stars are still growing, so the resulting BH masses may reach $\sim5\times10^4$--$10^5~{\rm M_\odot}$.
This separation is much larger than the $\sim3$--$5\times10^{-4}~{\rm pc}$ required for a circular binary to merge through gravitational-wave emission within a Hubble time.
Additional hardening through gas drag, dynamical friction, stellar scattering, or few-body interactions is therefore required before merger.
Nevertheless, the formation of a bound pair of massive BH progenitors demonstrates that the DCBH formation channel considered here can naturally provide initial conditions for high-redshift massive-BH mergers.

The SMSs formed in our simulations are expected to leave BHs with masses of $\sim10^4$--$10^5~{\rm M_\odot}$.
Mergers of heavy-seed BHs in this mass range may be detectable with future space-based gravitational-wave observatories such as the Laser Interferometer Space Antenna (LISA) and TianQin
\citep{Sesana+2007,Klein+2016,Wang+2019}.
Their mass and mass-ratio distributions may also help distinguish among different BH-seed formation scenarios
\citep{Sesana+2007,Klein+2016}.

High-redshift TDEs and gravitational-wave mergers may therefore provide complementary observational probes of the DCBH formation channel considered here.

\section{CONCLUSION}
\label{sec:conclusions}

In this study, we used cosmological radiation-hydrodynamic simulations
to follow the evolution of an atomic-cooling halo for
$\sim1$--$2~{\rm Myr}$
after the onset of star formation.
We focused on how internal radiative feedback from stars formed during
the initial star-formation phase affects the subsequent gas dynamics,
clustered star formation, and supermassive star (SMS) formation under subcritical external FUV backgrounds of $J_{21}=1$--$10^2$.

Our main results are summarized as follows.

\begin{itemize}

\item
In all runs, star formation in the atomic-cooling halo proceeds through
a characteristic two-stage evolution.
The gas first undergoes an initial collapse driven by H$_2$ cooling,
similar to ordinary Pop~III star formation.
Internal radiative feedback from the stars formed during this phase
then suppresses H$_2$ cooling and small-scale fragmentation, while the
surrounding atomic-cooling gas reservoir contracts under self-gravity
and undergoes global collapse.

\item
The global collapse involves a gravitationally bound gas reservoir extending over $\sim10$--$100~{\rm pc}$.
It leads to the formation of protostars with high accretion rates of $\sim0.1$--$1~{\rm M_\odot~yr^{-1}}$,
with the most massive objects reaching
$\sim10^4~{\rm M_\odot}$
by the end of the simulations.
Their accretion rates remain well above the critical value
$\simeq0.04~{\rm M_\odot~yr^{-1}}$
required to maintain the inflated supergiant-protostar state.
Since accretion is still ongoing, these objects are expected to grow
further toward
$\sim10^5~{\rm M_\odot}$
and become SMSs.

\item
The global collapse is not monolithic but is accompanied by clustered star formation.
Numerous stars with
$m_\star\lesssim10^3~{\rm M_\odot}$
form, but $\sim90\%$ of the stellar mass accreted during this phase is concentrated in a small number of rapidly accreting protostars with
$m_\star\gtrsim10^4~{\rm M_\odot}$.
The resulting stellar system can therefore be characterized as a strongly top-heavy system containing a few central SMSs surrounded by
a compact young stellar component.

\item
The SMSs formed through this pathway are expected to leave DCBHs with
masses of $\sim10^5~{\rm M_\odot}$,
which could grow further through accretion from the surrounding gas
reservoir.
The estimated abundance of these DCBHs, $n_{\rm DCBH}\sim10^{-3}$--$10^{-2}~{\rm cMpc^{-3}}$, is sufficiently high to provide a plausible progenitor population for
BH-powered LRDs and present-day SMBHs.

\item
The global collapse forms not only SMSs but also a compact young
stellar component concentrated within
$\sim1$--$10~{\rm pc}$.
Its stellar mass reaches
$\sim10^4~{\rm M_\odot}$
by the end of the simulations and could grow further with continued
star formation.
Its compactness, together with the expected UV and recombination-line
emission, suggests a possible connection to compact metal-poor and
Pop~III candidates observed with JWST.
The global-collapse scenario may therefore be relevant not only to
heavy-seed BH formation but also to the formation of compact stellar
systems in the early Universe.

\end{itemize}

These results show that, even under subcritical external FUV backgrounds, radiative feedback from stars formed earlier within an atomic-cooling halo can induce large-scale gas collapse and SMS formation.
The process simultaneously produces a small number of SMSs that dominate the stellar mass budget and a compact young stellar component capable of producing observable UV continuum and recombination-line emission.
The global-collapse scenario proposed here may therefore provide a common framework linking SMS formation, DCBH formation, high-redshift compact Pop~III systems, and the LRD population discovered with JWST.

Future simulations including supernova feedback, BH accretion feedback, metal enrichment, and longer-term evolution will be required to determine the final masses of the SMS progenitors,
the subsequent evolution of the stellar population after the global collapse, the mass supply onto the DCBHs, and their observational signatures.

\section*{ACKNOWLEDGEMENTS}
We would like to thank Kazumi Kashiyama, Takashi Hosokawa, Kazuyuki Sugimura, Daisuke Toyouchi, Kazutaka Kimura, Ryunosuke Maeda, and Tomoya Suzuguchi for fruitful discussions and helpful comments.
Numerical computations were carried out on the Cray XD2000 system Aterui II at the Center for Computational Astrophysics, National Astronomical Observatory of Japan. 
This work was supported by Grants-in-Aid for Scientific Research from the Ministry of Education, Culture, Sports, Science and Technology (MEXT) of Japan under Grant Nos. 22H00149 and 26H02060 (KO).
This work was supported by the Tohoku University NEXUS Research Fellow Program.
RS acknowledges support from the MUR projects FIS-2024-01621 DAWN and PRIN 2022CB3PJ3-FLAGS, the EU Recovery Fund PNRR, and the INFN TEONGRAV initiative.

%%%%%%%%%%%%%%%%%%%%%%%%%%%%%%%%%%%%%%%%%%%%%%%%%%
\section*{DATA AVAILABILITY}
The data underlying this article will be shared on a reasonable request of the corresponding author.



\bibliographystyle{mnras}
\bibliography{ms} % if your bibtex file is called example.bib

%%%%%%%%%%%%%%%%%%%%%%%%%%%%%%%%%%%%%%%%%%%%%%%%%%

%%%%%%%%%%%%%%%%% APPENDICES %%%%%%%%%%%%%%%%%%%%%
\appendix
\section{PHOTOIONIZATION FEEDBACK ON THE GLOBAL-COLLAPSE GAS RESERVOIR}
\label{app:photoionization_feedback}

In Section~\ref{sec:implications}, we discussed whether accretion onto the most massive stars can continue beyond the end of the simulations.
One potential limitation is photoionization feedback from the surrounding massive stars formed during the global collapse.
Their ionizing radiation could photoevaporate the dense gas reservoir and reduce the mass supply from larger scales.
Here, we estimate whether their ionizing photon emissivity is sufficient to ionize the surrounding dense envelope out to large radii.

We consider ionizing sources distributed within a characteristic radius $r_0$, surrounded by a dense gas envelope.
From the Larson--Penston-like envelope profile shown in
Figure~\ref{fig:r-nH-T-Mdot}, the gas density at $r>r_0$
can be approximated as
\begin{eqnarray}
n_{\rm H}(r)
&\simeq&
\frac{8.86 c_{\rm s}^2}{4\pi G m_{\rm H} r^2}
\nonumber\\
&\simeq&
5\times10^5~{\rm cm^{-3}}
\left(
\frac{T}{7\times10^3~{\rm K}}
\right)
\left(
\frac{r}{1~{\rm pc}}
\right)^{-2}.
\end{eqnarray}
For this density profile, we define the ionization radius
$r_{\rm ion}$ by balancing the ionizing photon emissivity $\Phi_{\rm ion}$ with the total recombination rate between $r_0$ and $r_{\rm ion}$:
\begin{eqnarray}
\Phi_{\rm ion}
&=&
\int_{r_0}^{r_{\rm ion}}
4\pi r^2 dr\,
\alpha_{\rm B} n_{\rm H}(r)^2
\nonumber\\
&=&
\Phi_{\rm crit}(r_0)
\left(
1-\frac{r_0}{r_{\rm ion}}
\right).
\end{eqnarray}
Here,
$\alpha_{\rm B}$
is the case-B recombination coefficient.
For an ionized-gas temperature of
$T_{\rm II}=10^4~{\rm K}$,
we adopt
$\alpha_{\rm B}=2.6\times10^{-13}~{\rm cm^3~s^{-1}}$.
The critical ionizing photon emissivity is
\begin{eqnarray}
\Phi_{\rm crit}(r_0)
&\equiv&
\frac{8.86^2\alpha_{\rm B} c_{\rm s}^4}
{4\pi G^2 m_{\rm H}^2 r_0}
\nonumber\\
&\simeq&
2.4\times10^{55}~{\rm s^{-1}}
\left(
\frac{r_0}{1~{\rm pc}}
\right)^{-1}
\left(
\frac{T}{7\times10^3~{\rm K}}
\right)^2.
\end{eqnarray}
This is the ionizing photon emissivity required to balance all
recombinations outside $r_0$ in an $r^{-2}$ envelope, corresponding to the limit $r_{\rm ion}\rightarrow\infty$.
For $\Phi_{\rm ion}<\Phi_{\rm crit}$,
the ionized region remains finite, with
\begin{eqnarray}
r_{\rm ion}
=
\frac{r_0}
{1-\Phi_{\rm ion}/\Phi_{\rm crit}(r_0)},
\end{eqnarray}
whereas no finite equilibrium ionization radius exists for
$\Phi_{\rm ion} \geq \Phi_{\rm crit}$.

We next consider a stellar population within a region of radius $\sim r_0$ consisting of $\sim10^3$ stars with masses of $\sim100~{\rm M_\odot}$, corresponding to a total stellar mass of $\sim10^5~{\rm M_\odot}$.
Assuming a luminosity of $\sim0.5L_{\rm Edd}$ and an effective temperature of $T_{\rm eff}\sim10^5~{\rm K}$ for each massive star, the total ionizing photon emissivity is approximately
\begin{eqnarray}
\Phi_{\rm ion}
\sim
10^{54}~{\rm s^{-1}}.
\end{eqnarray}
The ratio to the critical value is then
\begin{eqnarray}
\frac{\Phi_{\rm ion}}{\Phi_{\rm crit}}
&\simeq&
0.04
\left(
\frac{\Phi_{\rm ion}}
{10^{54}~{\rm s^{-1}}}
\right)
\left(
\frac{r_0}{1~{\rm pc}}
\right)
\left(
\frac{T}{7\times10^3~{\rm K}}
\right)^{-2}.
\end{eqnarray}
For $r_0\sim1~{\rm pc}$, we therefore obtain
$\Phi_{\rm ion}<\Phi_{\rm crit}$,
and the ionized region remains close to $r_0$.
For example, at
$r_0=1~{\rm pc}$,
\begin{eqnarray}
r_{\rm ion}
\simeq
1.04\,r_0.
\end{eqnarray}
The ionized region therefore extends only slightly beyond the stellar distribution. 

This estimate assumes static ionization equilibrium, a spherically
symmetric Larson--Penston-like envelope, and a spatially averaged
distribution of ionizing sources.
Nevertheless, the high recombination rate in the dense $r^{-2}$ envelope confines the ionized region close to the stellar distribution, making it difficult for ionizing radiation from the surrounding $\sim100~{\rm M_\odot}$ stars to affect the gas reservoir far beyond the pc scale.
Photoionization feedback from these stars is therefore unlikely to significantly suppress the gas supply from larger scales and terminate
accretion onto the most massive stars.
\label{lastpage}
\end{document}